\documentclass[aps,prc,tightenlines,twoside,
  fleqn,superscriptaddress]{revtex4-1}

\usepackage{amsmath,amsthm,amssymb}
\usepackage{graphicx}
\usepackage{soul}
\usepackage{color}
\usepackage{xcolor}

\colorlet{darkgreen}{green!50!black}
\colorlet{brightyellow}{yellow!75!red}
\colorlet{orange}{red!50!yellow}
\colorlet{darkblue}{blue!60!black}
\colorlet{darkred}{red!80!black}

\newcommand{\tfo}{T_{\mathrm{fo}}}
\newcommand{\mufo}{{\mathbf\mu}_{\mathrm{fo}}}
\newcommand{\taufo}{\tau_{\mathrm{fo}}}

\begin{document}

\title{Extraction of the Specific Shear Viscosity of Hot Hadron Gas}



\author{Zhidong Yang} 
\email{zdyang@comp.tamu.edu}

\author{Rainer J.\ Fries}
\email{rjfries@comp.tamu.edu}

\affiliation{Cyclotron Institute and Department of Physics and Astronomy, Texas A\&M University, College Station TX 77843, USA}

\date{\today}

\begin{abstract}
We extract the specific shear viscosity $\eta/s$ of nuclear matter for various temperatures and chemical potentials in the hadronic phase using data taken in high energy nuclear collisions.
We use a blastwave parameterization of the final state of nuclear collisions, including non-equilibrium deformations of particle distributions due to shear stress in the Navier-Stokes approximation. We fit spectra and elliptic flow of identified hadrons for a variety of collision energies and impact parameters at the Relativistic Heavy Ion Collider (RHIC) and the Large Hadron Collider (LHC). The systems analyzed cover a temperature range from about 110 to 140 MeV and vary in their chemical potentials for stable hadrons.
We attempt to assign meaningful systematic uncertainties to our results. This work is complementary to efforts using viscous fluid dynamics to extract the specific shear viscosity of quark gluon plasma at higher temperatures.
We put our work in context with existing theoretical calculations of the specific shear viscosity.
\end{abstract}


\maketitle

\section{Introduction}

Quark gluon plasma (QGP) and hot hadron gas are routinely created in nuclear collisions at the Relativistic Heavy Ion Collider (RHIC)
and the Large Hadron Collider (LHC).  Soon after the start of the RHIC program it was realized that experimental
data from this machine required a new paradigm. Although the original motivation for postulating the existence of quark gluon plasma 
came from the known weakening of the strong coupling constant with temperature \cite{Collins:1974ky,Shuryak:1978ij}, it turned out that 
QGP close to the pseudo-critical temperature $T_c$, i.e.\ at temperatures probed by the experimental programs at RHIC and LHC, rather behaves like a strongly coupled liquid \cite{Gyulassy:2004zy,Shuryak:2008eq}.
The first hint came from the great success enjoyed by \emph{ideal} fluid dynamics in describing the flow of hadrons measured at RHIC
\cite{Kolb:2000fha,Huovinen:2001cy}.
As it turns out, the process of cooling and expansion of the fireball of QGP and hadron gas behaves hydrodynamically from very early times
in the collision onward. Subsequently, relativistic \emph{viscous} fluid dynamic simulations compared to data allowed the quantitative 
extraction of $\eta/s$ from data \cite{Romatschke:2007mq,Dusling:2007gi,Schenke:2011tv,Heinz:2013th}. 
Around the same time, Kovtun, Son and Starinets hypothesized that there might be a universal lower bound of $\eta/s = 1/(4\pi)$ for the specific shear viscosity, based on their study of strongly interacting systems using AdS/CFT correspondence \cite{Kovtun:2004de}. Quark gluon plasma quickly became a celebrated example of an ideal liquid.

Measurements of the specific shear viscosity $\eta/s$ utilize viscous fluid dynamic simulations compared to experimental data. Collective flow observables are particularly sensitive to shear viscosity. 
The first generation of calculations used relativistic fluid dynamics with a fixed, temperature-independent
$\eta/s$ as a parameter. Fluid dynamics was run all the way to kinetic freeze-out in the hadronic phase 
which was modeled very similar to the approach discussed further below here. Obviously the value of $\eta/s$ extracted from
this method is averaged over the entire temperature evolution of the QGP \emph{and} the hot hadron gas below $T_c$, 
a range of several hundred MeV at top RHIC and LHC energies. $\eta/s$ extracted through this method also includes the effects of 
deformations of particle distributions at freeze-out that are present at finite shear stress \cite{Teaney:2003kp,Dusling:2007gi}.

Subsequently, several groups argued that the hadronic phase should be rather described by hadronic transport 
models because the specific shear viscosity in the hadronic phase could be too large for the evolution to be described 
accurately in second order viscous fluid dynamic codes \cite{Song:2010aq,Song:2011hk}. This argument was aided by estimates of 
$\eta/s$ for hadron gas from chiral perturbation theory, effective theories and hadronic transport by various groups 
\cite{Prakash:1993bt,Muroya:2004pu,Dobado:2003wr,Chen:2006iga,Itakura:2007mx,Dobado:2008vt,Demir:2008tr,FernandezFraile:2009mi,Romatschke:2014gna,Rose:2017bjz}. While these calculations do not agree quantitatively, they generally find rather large specific shear viscosity for a hot hadron gas, $\eta/s \gtrsim 4/(4\pi)$ even very close to $T_c$, see Fig.\ \ref{fig:final}.
Thus fluid dynamic calculations were matched to hadronic transport models just below $T_c$ while 
$\eta/s$ was retained as a parameter only for the QGP phase and the crossover region around $T_c$. Even more recently, fluid dynamic calculations using simple parameterizations have also been used to constrain the functional form of the temperature dependence of $\eta/s$,
mostly for the QGP case \cite{Song:2010aq,Niemi:2012ry,Bernhard:2015hxa}. 
We refer the reader to \cite{Gale:2013da,Heinz:2013th,Romatschke:2017ejr} for reviews of fluid dynamic simulations of nuclear collisions, including the extraction of shear viscosity.

Lattice calculations of $\eta/s$ in the QGP phase have been attempted but are challenging 
\cite{Meyer:2007ic,Meyer:2009jp,Haas:2013hpa,Mages:2015rea,Astrakhantsev:2017nrs}. They generally find $\eta/s$ to be close to the conjectured lower bound around $T_c$ with a rather slow rise towards higher temperatures. Perturbative QCD calculations at leading order have indicated large values of $\eta/s$ at temperatures well above $T_c$ \cite{Arnold:2000dr,Arnold:2003zc}, but a recent next-to-leading order calculation predicts a significant
drop towards $T_c$ which makes the perturbative results comparable to lattice QCD \cite{Ghiglieri:2018dib}. 

From general arguments one expects a minimum of $\eta/s$ around $T_c$ which has been found to be the case for a large variety of systems \cite{Csernai:2006zz}. How fast the specific shear viscosity is rising towards lower temperatures below $T_c$ can not be seen as settled from either data nor first principle calculations. Features should be continuous as 
a function of temperature but could be changing quickly. Progress has been made in undestanding the
large values of $\eta/s$ in hadronic transport \cite{Rose:2017bjz}. However, these effective theories mostly do not incorporate the existence of a pseudo-critical temperature and their predictions close to this temperature should be 
viewed with caution. The question of the specific shear viscosity of hadron gas is an important one. In any conceivable experiment information on specific shear viscosity in the QGP phase is always convoluted with contributions from the hadronic phase. Thus uncertainties in hadronic 
$\eta/s$ are directly responsible for increased uncertainties of QGP shear viscosities extracted from data.

It is clear that an independent assessment of the hadronic specific shear viscosity is necessary to improve the situation. As a reasonable minimum requirement, theoretical uncertainties coming from incomplete knowledge 
of the hadronic phase should inform realistic contributions to error bars for quantities extracted for the QGP phase.
For this reason we attempt a realistic uncertainty estimate in this work,
Moreover, the specific shear viscosity of a hot hadron gas is by itself a compelling question.

Here, we argue that it is possible to use experimental data to estimate the specific shear viscosity of the hot hadron
gas at the kinetic freeze-out independently. The main effect of the time evolution of the system before freeze-out is the build-up of a flow 
field $u^\mu$ which leads to the system expanding and cooling. Viscous corrections to first order are given by gradients of the flow field 
(Navier-Stokes approximation). Computing the flow field in fluid dynamics, introduces additional dependences on initial conditions 
and the equation of state. We take a complementary approach and \emph{fit} the final flow field, together with the temperature and 
system size at kinetic freeze-out. The specific shear viscosity is then a parameter at just one fixed temperature $T=\tfo$, the kinetic freeze-out temperature, and a set of chemical potentials $\mufo=(\mu_B,\mu_\pi,\ldots )$
for baryon number $B$, and abundances of stable hadrons like pions, kaons and nucleons. Of course, such fits of flow fields and temperatures at freeze-out are well established and generally known as blastwave parameterizations \cite{Westfall:1976fu,Schnedermann:1993ws,Retiere:2003kf}. We will use such a blastwave, with $\eta/s$ added as a parameter, to extract $\eta/s(\tfo,\mufo)$ for a variety of points ($\tfo$,$\mufo$) in different collision systems. Here we present the results for Au+Au collisions at top RHIC energies and Pb+Pb collisions at LHC
where the baryon chemical potential vanishes $\mu_B \approx 0$. However, non-vanishing chemical potentials
$\mu_\pi$, $\mu_K$ and $\mu_p$ are present at kinetic freeze-out, determined by the chemical freeze-out at
higher temperatures.
Our extraction is complementary to fluid dynamics, which integrates over the effects of shear viscosity over a wide temperature range. Some of the 
uncertainties in both approaches are the same. For example the assumption of a sharp kinetic freeze-out at a fixed temperature is common to 
both approaches and is only an approximation, although it can be improved in the case of fluid dynamics by matching to hadronic transport. Other uncertainties are different in both approaches. For example the dependence of fluid
dynamic calculations on initial conditions, which themselves are not well constrained experimentally, is not present in our approach. We will
discuss uncertainties in the blastwave extraction in more detail below. This paper is organized as follows: In section 2 we present our blastwave parameterization, especially the viscous correction term. In section 3 we discuss the simulation and data selection. In section 4 we show fit results and carry out an analysis of uncertainties. We conclude with a discussion
and outlook in section 5.

\section{A Viscous Blast Wave}
\label{sec:2}

Viscous corrections to blastwaves have been studied in \cite{Teaney:2003kp,Jaiswal:2015saa}. Both of these previous works assume spatial spherical symmetry in the transverse plane and free streaming for simplicity. We will generalize these assumptions here. We choose the blastwave of Retiere and Lisa (RL) \cite{Retiere:2003kf} as our starting point. In this section we discuss the Retiere-Lisa blastwave and compute the Navier-Stokes corrections.

We have to make two major assumptions in our analysis, both of which have been routinely used and studied in the literature. The first is that at freeze-out the system of hadrons is close enough to kinetic equilibrium so that at any position $r^\mu=(t,x,y,z)$ there exist a local rest frame with a local temperature $\tfo(r)$ and a set of chemical potentials $\mufo(r)$ 
such that the particle distribution in the local rest frame can be written as
\begin{equation}
  \label{eq:fdist}
  f(r,p) = f_{0}(r,p) + \delta f(r,p)
\end{equation}
where $f_0$ is the equilibrium Bose/Fermi-distribution with the local temperature and chemical potentials, 
\begin{equation}
  f_0(r,p) = \frac{1}{e^{(E-\mu(r,p))/\tfo(r,p)}\mp 1}  \, , 
\end{equation}
and $\delta f$ is a gradient correction of Navier-Stokes type. Here we use the general form 
\begin{equation}
  \delta f (r,p) = \frac{\eta}{s} \frac{\Gamma(6)}{\Gamma(4+\lambda)} \left(\frac{E}{\tfo}\right)^{2-\lambda} 
  \frac{p^\mu p^\nu}{\tfo^3}\sigma^{\mu\nu} f_{0}(r,p)
\end{equation}
which follows from a generalized Grad ansatz \cite{Damodaran:2017ior}. In the Navier-Stokes approximation the viscous correction is proportional to the traceless shear gradient tensor, defined as
\begin{equation}
  \sigma^{\mu\nu} = \frac{1}{2} \left( \nabla^\mu u^\nu + \nabla^\nu u^\mu \right) - \frac{1}{3} \Delta^{\mu\nu} \nabla_\lambda u^\lambda  \, .
\end{equation}
Here $\nabla^\mu = \Delta^{\mu\nu} \partial_\nu$, with $\Delta^{\mu\nu} = g^{\mu\nu} - u^\mu u^\nu$, is the derivative perpendicular to the flow field vector $u^\mu$. The gradient corrections need to be small and we will ensure that numerically $\delta f \lesssim f$ for all
relevant momenta in this analysis. The power $\lambda$ parameterizes further details of the underlying microscopic physics. Here we restrict ourselves to the original Grad ansatz $\lambda=2$ which is widely used. 
We reserve a more detailed analysis including $\lambda$ as a tunable parameter for future work. 

The second major assumption in our analysis pertains to the simplified shape of the freeze-out hypersurface and flow field. 
In longitudinal direction (along the colliding beams) we assume boost-invariance, which is a good approximation for particles measured around midrapidity at LHC and top RHIC energies. Blast wave parameterizations assume that
freeze-out happens at constant $\tfo$ and $\mufo$ which is approximated by a constant (longitudinal) proper time $\tau=\taufo$.
In the RL parameterization the transverse shape of the fireball at freeze-out is assumed to be an ellipse with semi axes $R_x$ and $R_y$ in $x$- and $y$-directions respectively. We define the coordinate axes such that the impact parameter $b$ of the collision is measured along the $x$-axis. In the following we use the reduced radius $\rho=\sqrt{x^2/R_x^2+y^2/R_y^2}$.
The flow field can be parameterized as 
\begin{equation}
  \label{eq:upara}
  u^\mu = \left( \cosh \eta_s \cosh \eta_T, \sinh \eta_T \cos\phi_u, \sinh\eta_T \sin\phi_u, \sinh\eta_s \cosh\eta_T
  \right)
\end{equation}
where $\eta_T$ is the transverse rapidity in the $x-y$-plane, and $\phi_u$ is the azimuthal angle of the flow
vector in the transverse plane. Boost invariance fixes the longitudinal flow rapidity to be equal
to the space time rapidity $\eta_s = 1/2 \log[(t+z)/(t-z)]$. For the transverse flow velocity $v_T = \tanh \eta_T$ we make the assumption
\cite{Retiere:2003kf}
\begin{equation}
  v_T =  \rho^n \left(\alpha_0 + \alpha_2 \cos(2 \phi_u) \right)
\end{equation}
which encodes a Hubble-like velocity ordering with an additional shape parameter $n$. $\alpha_0$ is the average velocity on the 
boundary $\rho=1$, and $\alpha_2$ parameterizes an elliptic deformation of the flow field coming from the original elliptic spatial deformation of systems with finite impact parameters. The time evolution of pressure gradients in the expansion leads to
flow vectors tilted towards the smaller axis of the ellipse. This is accomplished by demanding that the transverse flow vector is perpendicular to the elliptic surface at $\rho=1$,  i.e.\ $\tan \phi_u = R_x^2/R_y^2 \tan \phi$, where $\phi=\arctan y/x$ is the azimuthal angle of the position $r^\mu$. 
Higher order deformations could be present \cite{Jaiswal:2015saa}, but the two main observables chosen for our analysis are not particularly sensitive to them.

Using the assumptions laid out here we can write the spectrum of hadrons emitted from freeze-out as \cite{Cooper:1974mv}
\begin{equation}
  \frac{dN}{dy d^2P_T} = g \int\frac{p\cdot d\Sigma}{(2\pi)^3} f(r, u\cdot p)
\end{equation}
where $g$ is the degeneracy factor for a given hadron. The momentum vector in the laboratory frame is written in
standard form as 
$p^\mu = (M_T \cosh y, P_T \cos\psi,$ $P_T \sin\psi, M_T\sinh y)$ in terms of the transverse momentum $P_T$, the 
longitudinal momentum rapidity $y$ and the azimuthal angle $\psi$ in the transverse plane. $M_T^2 = P_T^2+M^2$ 
defines the transverse mass $M_T$ for a hadron of mass $M$. $d\Sigma^\mu$ is a parameterization of the $T=\tfo$ 
hypersurface and its outbound normal vector. With our assumptions we have $d\Sigma^\mu = \taufo R_x R_y d\eta_s \rho d\rho d\theta
(\cosh\eta_s,0,0,\sinh\eta_s)$. Hence, for hadrons measured around midrapidity ($y=0$) the spectrum takes the standard form
\begin{equation}
  \frac{dN}{dy d^2P_T} = g \taufo R_x R_y M_T \int_0^1 d\rho \int_0^{2\pi} d\theta \int_{-\infty}^\infty d\eta_s \frac{\rho \cosh\eta_s}{(2\pi)^3}  
   f_0(\rho,\theta,\eta_s; u\cdot p) \left[ 1+\frac{\eta}{s} \frac{1}{\tfo^3} p^\mu p^\nu \sigma^{\mu\nu} \right]  
  \label{eq:spectrum}
\end{equation}
where $\tan\theta = R_x/R_y \tan\phi$.
The set of parameters in this ansatz is $\tilde{\mathcal{P}} = (\taufo, R_x, R_y,\tfo,\mufo, n, \alpha_0, \alpha_2, \eta/s)$.

We can now determine the shear gradient tensor $\sigma^{\mu\nu}$ for the RL blastwave, following the example of
\cite{Teaney:2003kp,Jaiswal:2015saa}.
Without azimuthal symmetry the spatial derivatives in $\sigma^{\mu\nu}$ are still straight forward to obtain, starting from the explicit expression in Eq.\ (\ref{eq:upara}), but the results are somewhat lengthy. We delegate a discussion of details of these expressions to another work. 
The task of determining the time-derivatives in $\sigma^{\mu\nu}$ can be reduced to the question of computing $\partial_\tau \eta_T$ and 
$\partial_\tau \phi_u$. We start from the relativistic fluid dynamic equations of motion $\partial_\mu T^{\mu\nu}=0$, where $T^{\mu\nu}=e u^\mu u^\nu - p \Delta^{\mu\nu}$ is the ideal energy momentum tensor. We can restrict ourselves to ideal fluid dynamics to obtain the leading order expressions in a gradient expansion for the time derivatives. Dissipative corrections in the determination of the time derivatives would 
lead to terms of order $\eta^2 \times \text{(second order spatial gradients)}$ in $\delta f$ which we neglect. Here $e$ is the local energy density and $p$ the pressure.
The ideal fluid dynamics equations can be rewritten more instructively as the set of equations
\begin{align}
  \label{eq:fd1}
  De=& -(e+p)  \partial_\mu u^\mu=-(e+p)\nabla_\mu u^\mu
  \\
  \label{eq:fd2}
  D u^\mu=& \frac{\nabla^\mu p}{e+p}
\end{align}
where the co-moving time derivative is $D = u_\mu \partial^\mu$.

Freeze-out is the process of decoupling of particles where the mean free path rapidly grows beyond the system size. In fluid dynamics this
process is modeled through a sudden transition during which the mean free path goes from very small values to infinity instantaneously 
at $T = \tfo$. The system is free streaming, $D u^\mu = 0$, after the transition, i.e.\ from $T=\tfo-\epsilon$ on (with small $\epsilon > 0$). 
Thus the assumption of free streaming has been used in some previous work on viscous blastwaves \cite{Teaney:2003kp,Jaiswal:2015saa}. 
However, it seems more physical to assume that the local particle distributions $f(r,p)$ remain frozen across the $T=\tfo$ hypersurface and that 
$\sigma^{\mu\nu}$, including time derivatives, should be set at temperature $T=\tfo+\epsilon$. This is consistent with the treatment in fluid dynamics. Eqs.\ (\ref{eq:fd1}), (\ref{eq:fd2}) can be solved for the blast wave geometry and flow field assumed here to obtain the time derivatives we seek.

Using (\ref{eq:fd1}) together with the first or fourth equation ($\mu=0, 3$; the two equations are equivalent) in (\ref{eq:fd2}) we obtain
the time derivative of the transverse flow rapidity,
\begin{equation}
  \label{eq:etaTderivative}
  (1- c_s^2 \tanh^2\eta_T ) \partial_\tau \cosh\eta_T   =c_s^2 \tanh^2\eta_T (\partial_1u^1+\partial_2u^2+  \frac{\cosh\eta_T}{\tau})-\frac{u^1\partial_1u^0}{u^0}-\frac{u^2\partial_2u^0}{u^0}  \, ,
\end{equation}
in terms of known spatial derivatives. $c_s^2 = \partial p /\partial e$ is the speed of sound squared, given by the equation of state of the system at 
$T=\tfo$. The time derivative of the direction of the transverse flow field can be computed by using (\ref{eq:etaTderivative}) in the second
and third equation ($\mu=1, 2$) in (\ref{eq:fd2}).

This completes the brief introduction of the blastwave model used in our analysis. We can in principle
validate the blastwave by comparison with established fluid dynamics calculations. 
Here we will use the viscous fluid code MUSIC \cite{Schenke:2010nt,Ryu:2015vwa} for this purpose. 
Since the blastwave, for simplicity, ignores the effects of feed-down from hadronic resonances as well as 
bulk stress, we can quantify the uncertainty and bias from these simplifications by comparing to MUSIC calculations \emph{with} resonance decays and bulk stress included. This comparison is important to estimate uncertainties in the extraction of $\eta/s$ and will be discussed in some detail below.

\begin{table}[tb]
\begin{tabular}{|c|c|c|c|c|c|c|}
\hline
Centrality & proton (GeV/$c$) & kaon (GeV/$c$) & pion (GeV/$c$) &  $b$ (fm) & $c_s^2$ & $c_\tau$  \\
 \hline
\hline\hline
\multicolumn{7}{|l|}{ALICE 2.76 TeV }  \\
 \hline\hline
10-20\% & 0.325-3.3 & 0.225-2.55 & 0.525-1.85 & 6.05 &  0.158 & 0.783 \\
 \hline
20-30\% & 0.325-3.1 & 0.225-2.35 & 0.525-1.75 & 7.81 &  0.162 & 0.756 \\
 \hline
30-40\% & 0.325-3.1 & 0.225-2.25 & 0.525-1.65 & 9.23 &  0.166 & 0.720 \\
 \hline
40-50\% & 0.325-2.95 & 0.225-2.15 & 0.525-1.45 & 10.47 &  0.170 & 0.679 \\
 \hline
50-60\% & 0.325-2.55 & 0.225-1.85 & 0.525-1.25 & 11.58 &  0.174 & 0.633 \\
\hline\hline
\multicolumn{7}{|l|}{PHENIX 0.2 TeV} \\
 \hline\hline
 10-20\% & 0.55-2.9 & 0.55-1.85 & 0.55-1.65 & 5.70 &  0.164 & 0.780 \\
 \hline
 20-40\% & 0.55-2.7 & 0.55-1.75 & 0.55-1.55 & 8.10 (7.4, 8.7) &  0.170 & 0.739 \\
 \hline
 40-60\% & 0.55-2.5 & 0.55-1.65 & 0.55-1.45 & 10.5 (9.9, 11,0) &  0.178 & 0.660 \\
  \hline\hline
  \multicolumn{7}{|l|}{MUSIC 0.2 TeV} \\
   \hline\hline
  spectra ($v_2$)  & 0.25-2.34 (3.0) & 0.31-2.0 (3.0) & 0.38-1.77 (3.0) & 7.5 &  0.178 & 0.60 \\
  \hline
\end{tabular}
\caption{\label{table:allrange} Regular fit range (RFR) selected for each ALICE and PHENIX centrality bin for the spectra of all three particle species. The bins for elliptic flow data are chosen consistently.
We also show the average impact parameter $b$ from Glauber Monte Carlo calculations quoted by the experiments, 
the speed of sound squared $c_s^2$ and the expansion parameter $c_\tau$ determined for each data set.
For PHENIX data the average impact parameter for the two 10\% bins included in a given 20 \% bin are quoted in parentheses.
The bottom part of the table shows out choice of fit range for the spectra obtained from the MUSIC hydro code, see Sec.\ \ref{sec:4}.
MUSIC $v_2$ output is fit up to 3 GeV/$c$.}
\end{table}

\section{Simulation and Data Selection}
\label{sec:3}

With $\sigma^{\mu\nu}$ known it is straightforward to evaluate Eq.\ (\ref{eq:spectrum}) numerically, dependent on the set of parameters $\tilde{ \mathcal{P}}$ which can be determined from the fits to data or other methods.
We carry out this analysis using data on identified protons and antiprotons, kaons and pions from LHC and RHIC. We utilize both transverse momentum spectra around mid-rapidity, and elliptic flow $v_2$, the leading harmonic deformation of the spectrum in azimuthal momentum space angle $\psi$,
as functions of hadron transverse momentum $P_T$. They are calculated from (\ref{eq:spectrum}) as
\begin{align}
   \frac{dN}{2\pi P_T dP_T dy} &= \frac{1}{2\pi} \int d\psi \frac{dN}{dy d^2 P_T}   \, ,   \\
   v_2 (P_T) &= {\left(   \frac{dN}{2\pi P_T dP_T dy}  \right)}^{-1} \frac{1}{2\pi} \int d\psi \cos(2\psi) \frac{dN}{dy d^2 P_T}   \, ,
\end{align}
respectively. Note that the blastwave does not incorporate fluctuations. This is one reason why we will not analyze
the most central and peripheral centrality bins available which are known to exhibit large effects due to fluctuations.
All expressions in the blast wave are taken at rapidity $y=0$ and we have utilized matching data sets that have been taken around midrapidity.

We use data from the ALICE collaboration for Pb+Pb collisions at 2.76 TeV \cite{Adam:2015kca,Abelev:2014pua}, in 10\% centrality bins, and from the PHENIX collaboration for Au+Au collisions at 200 GeV \cite{Adare:2013esx,Adare:2014kci}. 
The PHENIX data is binned in 10 or 20\% centrality bins for the spectra and 10\% centrality bins for elliptic flow. 
For this analysis, if the PHENIX spectrum is only availabe in a coarser bin we combine a given 10\% bin for elliptic flow together with the overlapping 20\% bin for the spectrum. We find that centralities that share the coarser spectrum bins give 
results for temperature and specific shear viscosity that agree very well with each other within estimated uncertainties.

The selection of data points for the fit can introduce a bias that we try to quantify as an uncertainty. The following general principles were applied in the selection. 
We expect the blastwave parameterization to extract inaccurate parameters at too low momenta where resonance decays dominate the spectrum \cite{Sollfrank:1991xm}. We also expect it to fail at too large momentum where gradient corrections become large, and hadrons from other production channels, like hard processes, start to dominate soft particles from the bulk of the fireball.  The maximum momentum $P_T$  described by the blastwave increases from peripheral to more central collisions, since particles are expected to be more thermalized when volumes and lifetimes are larger. In addition, flow pushes particles with the same velocity to higher momentum if their mass is larger. Thus fit ranges for heavier particles can extend farther. 

Using these guiding principles we choose a preferred fit range in transverse momentum for each centrality, collision energy and particle species. We call this selection the regular fit range (RFR). For example, the regular fit range for the ALICE data in the 30-40\% centrality bin uses data points for the spectra in the $P_T$-intervals 0.325-3.10 GeV/$c$, 0.225-2.25 GeV/$c$ and 0.525-1.65 GeV/$c$ for protons, kaons and pions, respectively. The RFR for all data sets used here is shown in Tab.\ \ref{table:allrange}.
The $v_2$ data points included in this analysis are chosen to be consistent with the spectrum data points.
We note that our fit ranges for ALICE data extend to higher momentum compared to the fit ranges previously used by the ALICE collaboration for their blastwave fits without viscous corrections \cite{Abelev:2013vea}.
For each data set we supplement the regular fit ranges with lower (LFR) and higher (HFR) fit ranges in an attempt to quantify uncertainties from fit range selection. This will be discussed in detail in the next section.

\begin{table}[tb]
\begin{tabular}{|l|l|l|l|l|}
 \hline
centrality & $\mu_\pi$ (MeV) & $\mu_K$ (MeV) & $\mu_p$ (MeV) &  $T$ (MeV)  \\
\hline\hline
\multicolumn{5}{|l|}{ALICE 2.76 TeV }  \\
 \hline\hline
10-20\% & 70 & 100 & 245 & 113 \\
 \hline
20-30\% & 64 & 85 & 220 & 118 \\
 \hline
30-40\% & 61 & 73 & 203 & 121 \\
 \hline
40-50\% & 58 & 63 & 190 & 126 \\
 \hline
50-60\% & 55 & 47 & 170 & 130 \\
\hline\hline
\multicolumn{5}{|l|}{PHENIX 0.2 TeV }  \\
 \hline\hline
10-20\% & 65 & 62 & 200 & 121 \\
 \hline
20-40\% & 61 & 51 & 188 & 124 \\
 \hline
40-60\% & 53 & 22 & 138 & 134 \\
\hline
\end{tabular}
\caption{Chemical potentials for pion, kaon and proton for each ALICE and PHENIX data set in its regular fit range, together
with the extracted freeze-out temperatures.
\label{table:chem}}
\end{table}

\begin{figure}[tb]
\centering
\includegraphics[height=5 in,width=\textwidth]{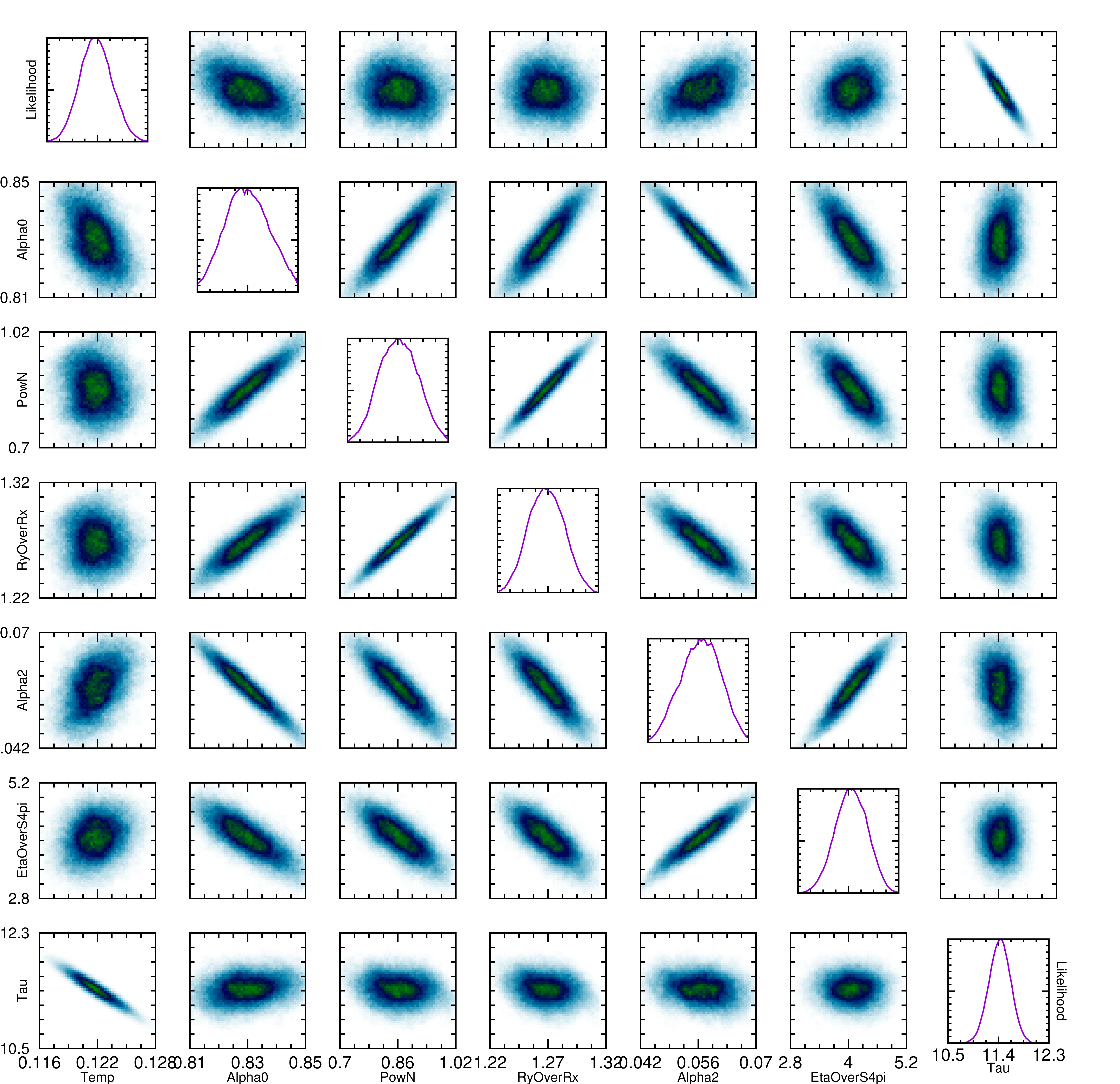}
\caption{Likelihood analysis for ALICE data in the 30-40 $\%$ centrality bin provided by the MADAI package. The horizontal and 
vertical axes show the chosen prior ranges for the parameters $\mathcal{P}$. From left to right (bottom to top): 
$T$ (MeV), $\alpha_0$, $n$, $R_y/R_x$, $\alpha_2$, $\eta/s$, $\tau$ (fm/$c$).
Plots on the diagonal show posterior likelihood distributions. The off-diagonal plots show correlations between parameters.
\label{fig:madai}}
\end{figure}

We use the statistical analysis package from the Models and Data Analysis Initiative (MADAI) project \cite{MADAI 2013,Bass 2016} to determine fit parameters. The MADAI package includes a Gaussian process emulator and a Bayesian analysis tool. A single computation of Eq.\ (\ref{eq:spectrum}) is quite fast. The Gaussian process emulator allows us to carry out the full statistical analysis easily on a single CPU. We choose appropriate prior ranges for each parameter (see Fig.\ \ref{fig:madai} for an example) with flat probabilities within each range. We use 500 training points for the Gaussian process emulator (800 for the 10-20\% PHENIX bin). We check that the results of the Gaussian emulator are within a few percent of the true blast wave result. Finally a Markov Chain Monte Carlo provides a likelihood analysis and gives the maximum likelihood parameters and uncertainties.

As discussed above, for the analysis here we will set $\lambda=2$. We will further restrict the set of simultaneously fitted parameters to seven, choosing $\mathcal{P}=(\taufo, \tfo, R_y/R_x, n, \alpha_0, \alpha_2, \eta/s)$ from the full set $\tilde{ \mathcal{P}}$.
Two considerations guide our choice to restrict the number of parameters. Some of the parameters we have removed are highly correlated with remaining ones. Sometimes the correlation can be more easily resolved by additional theoretical considerations. 
For example, our chosen observables depend on 
$R_x$, $R_y$ and $\taufo$ primarily through the ratio $R_y/R_x$, which is a main driver for elliptic flow, and through the overall volume $\sim R_xR_y\taufo$ which determines the normalization of spectra. Dependences on the individual size parameters are
absent in the ideal blast wave, but enter in a sub-leading way through the viscous correction terms.
We constrain $R_x$, $R_y$ and $\tfo$ by fitting the ratio $R_y/R_x$, and the time $\taufo$ and by adding in addition the simple 
geometric estimate
\begin{equation}
  R_x \approx (R_0-b/2)+ \taufo c_\tau (\alpha_0+\alpha_2)  \, , \\ 
\end{equation}
for the propagation of the fireball boundary in $x$-direction.
Here the radius of the colliding nucleus is $R_0$, and the impact parameter is denoted as $b$. The expansion parameter
$c_\tau = \bar\alpha_0 / \alpha_0$ relates the \emph{time-averaged} surface velocity $\bar\alpha_0$ with its final value $\alpha_0$
at freeze-out. The boundary velocity parameters $\alpha_0$ and $\alpha_2$ at freeze out are fitted to data. 
$c_\tau$ can be estimated to be between 0.6 and 0.8 going from the most peripheral bin to the most central bin in the analysis. This can be inferred from typical radial velocity-vs-time curves obtained in fluid dynamic simulations \cite{Song:2009rh}. 
As this is a simple model we vary $c_\tau$ in the next section to explore the uncertainties from this choice of parameter reduction. 
The impact parameter $b$ used for each centrality bin is taken from Glauber Monte 
Carlo simulations used by the corresponding experiment \cite{Abelev:2013vea,Adler:2003qi}.

The speed of sound squared $c_s^2$ for a hadronic gas is discussed e.g.\ in \cite{Teaney:2002aj,Huovinen:2010cs2}.
We use \cite{Teaney:2002aj} to adjust $c_s^2$ iteratively with the temperature found for each fitted centrality and collision
system. The values we find are given in Tab.\ \ref{table:allrange} for quick reference. Further below we will explore
the dependence of the extracted shear viscosity and temperature on our choice of speed of sound by varying $c_s^2$.
The relevant chemical potentials are not quite settled in the literature. We find good fits for chemical potentials for pions
roughly consistent with \cite{Hirano:2002qj,Teaney:2002aj}. The values for ($\mu_\pi,\mu_K,\mu_p)$ for each data set 
are summarized in Tab.\ \ref{table:chem}. Again we account for the uncertainties by varying the chose values 
in the uncertainty analysis in the next section.

\begin{table}

\begin{tabular}{|l|l|l|l|l|l|}
 \hline
Total error & ALICE & \multicolumn{3}{|c|}{PHENIX} &  MUSIC  \\
\hline
  & 30-40\% & 10-20\% & 20-40\% &40-60\% &   \\
 \hline
spectra(\%) & 5.65 & 1.23 & 0.89 & 0.92& 4.0 \\
 \hline
$v_2$(\%) & 3.24 & 6.71 & 3.13, 3.29 &3.27, 3.80& 2.0 \\
\hline
\end{tabular}
\caption{Typical error percentage, defined as the median for all bins in the RFR, for PHENIX data. The statistical error only is shown for
the spectra. For comparison we also show one centrality bin of ALICE data. When two values for the error on $v_2$ are given they refer to the values in the smaller 10\%-wide centrality bins covered. We also show the error assumed for the fit to fluid dynamic simulation (MUSIC), see text for details.
\label{table:error}}
\end{table}

Error bars for experimental data are crucial input for the statistical analysis. In absence of further details about correlations between
error bars we use the statistical and systematic errors quoted by experiments, summed in quadrature, for each momentum bin. This
is the main uncertainty input to the MADAI analysis. This procedure works well for ALICE data.
Systematic errors for PHENIX identified hadron $P_T$-spectra are discussed in \cite{Adare:2013esx} but numbers are not included in the published data files. We thus start with the provided statistical errors and scale them up. Interestingly, the statistical analysis itself
also suggests that statistical error bars alone for the PHENIX $P_T$-spectra are insufficient in the presence of much larger uncertainties for elliptic flow. This comes about because there is a competition between fits to $P_T$-spectra and $v_2$ regarding the best value of $\eta/s$. Momentum spectra prefer small viscous corrections, while $v_2$ data typically prefers large viscous corrections. The optimized $\eta/s$ will be a balance between these constraints. 
If error bars are unbalanced between spectra and $v_2$ we see large likelihoods but nevertheless ill-fitting approximations for the quantity with larger error bars. We have to assume that the extraction of $\eta/s$ is then biased in one direction. 
It is suggestive to accept an overall larger uncertainty for possibly less bias in the analysis.
As a result of these considerations we multiply the statistical error given for PHENIX spectra by factors of 1.5, 3 and 4 for the 10-20\%, 20-40\% and 40-60\% centrality bins, respectively. Similar considerations apply for the fit to fluid dynamic simulations discussed below. Table \ref{table:error} shows the typical relative error in some data sets in the regular fit range (RFR), before adjustments are made. The typical value is defined as the median value within the RFR for all three hadron species.

\section{Fit Results and Uncertainty Analysis}
\label{sec:4}

With the preparations from the previous sections in place we go ahead and analyze the available data for
each energy and centrality bin. The fit results are generally of good quality despite the relatively large RFR fit range.
As an example we discuss here the 30-40\% centrality bin for ALICE data in detail. Fig.\ \ref{fig:madai} shows the results for the 
fit parameter set $\mathcal{P}$ from the statistical analysis for fits in the RFR of this data set. The likelihood plots on the diagonal of Fig.\ \ref{fig:madai} show well defined peaks. The off-diagonal plots show correlations between fit parameters. 
The preferred (average) values for this ALICE centrality bin are $\tau=11.41$ fm/$c$, $T$=121.9 MeV, $\alpha_0$=0.830$c$, $n$=0.87, $R_y/R_x$=1.270, $\alpha_2$=0.0564$c$, $\eta/s$=4.06/4$\pi$. Recall that the values for the external parameters 
$c_\tau$ and $c_s^2$ as well as for the chemical potentials, and the regular fit range used are given in Tab.\ \ref{table:allrange} and \ref{table:chem}, respectively.

Although we have already eliminated some parameters from the blastwave there are still correlations between the remaining parameters
in $\mathcal{P}$. Most prominently there is an expected anti-correlation between freeze-out time and temperature which 
comes from the constraint on the overall number of particles. Surprisingly there is no pronounced anti-correlation between temperature and radial flow parameter $\alpha_0$, which means that the choice of three different hadrons to fit, and the sizes of the fit ranges, are sufficient to cleanly separate thermal and collective motion. We note a correlation between the elliptic flow parameter $\alpha_2$ and $\eta/s$. As expected, for larger values of $P_T$ these two parameters move the elliptic flow in different directions, i.e.\ an increase in one of these parameters will necessitate an increase in the other one. The correlations seen in this centrality bin are found to be qualitatively true for the other energies and centrality bins as well.

\begin{figure}[tb]
\centering
\includegraphics[width=3.5 in]{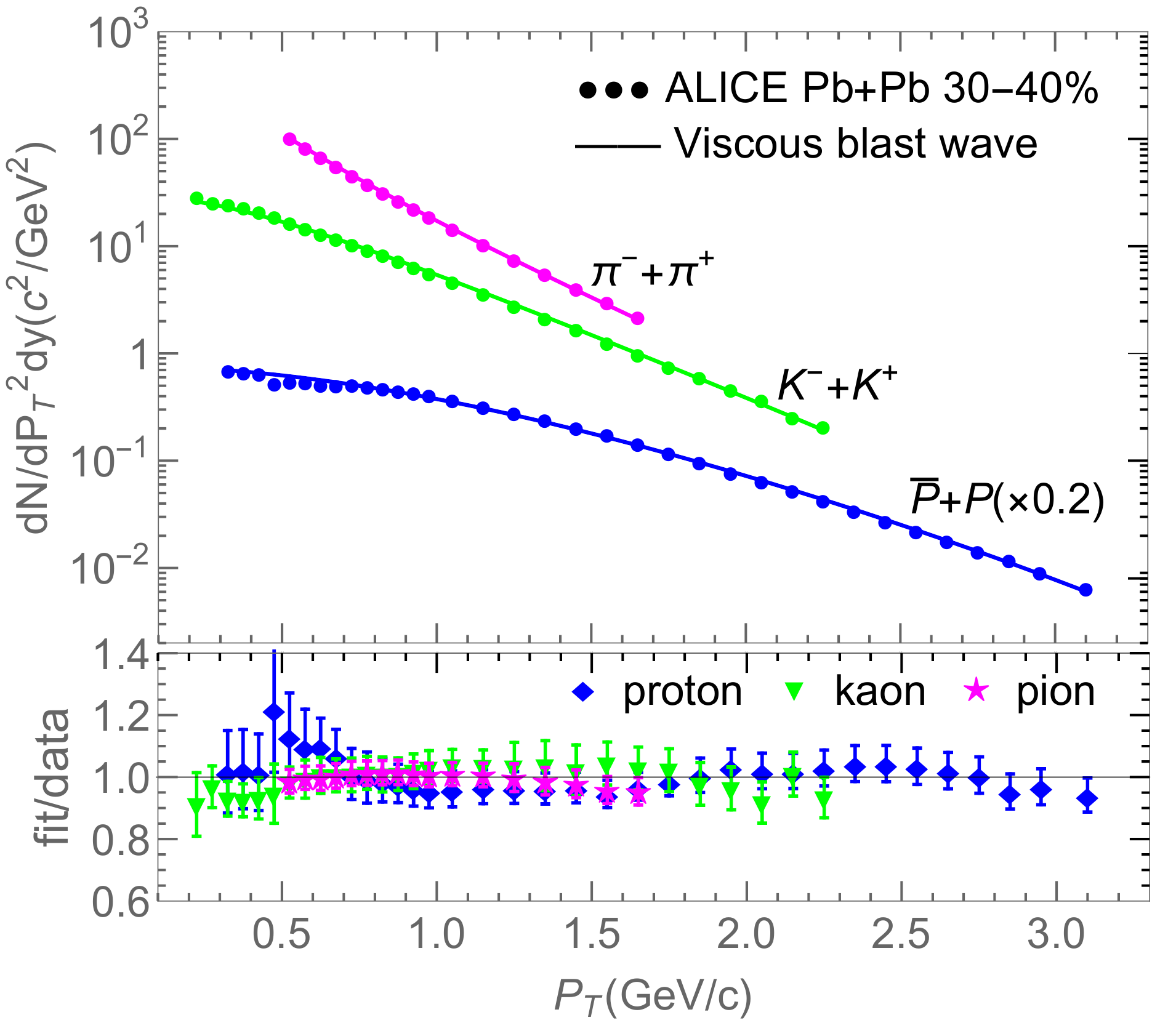}
\includegraphics[width=3.5 in]{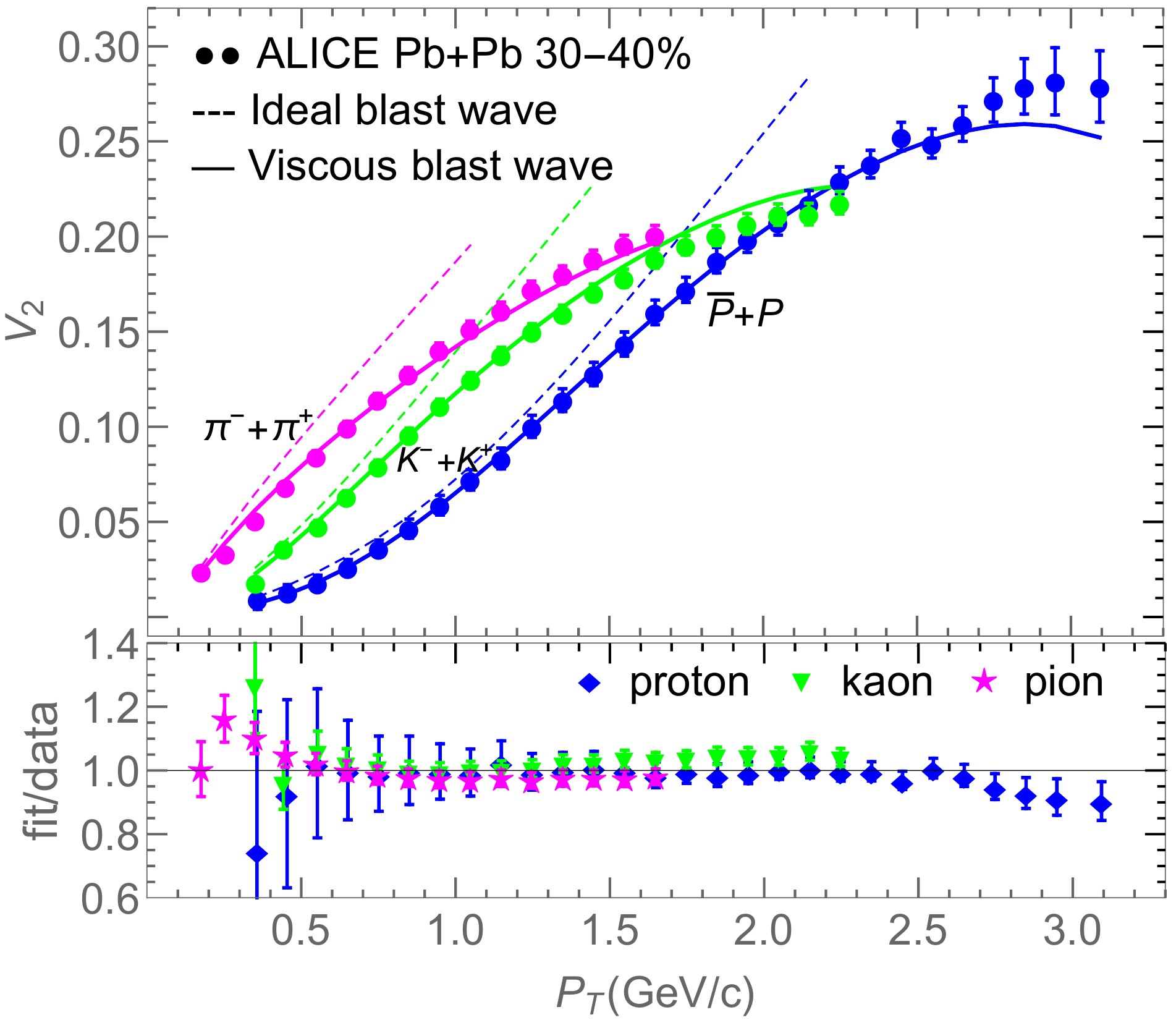}
\caption{\label{fig:3040data}
Left panel: Transverse momentum spectra for pions, kaons and protons (solid lines), respectively, using the extracted, preferred fit parameters 
for the ALICE 30-40 $\%$ centrality bin. We also show the ALICE data used for the fit (circles) with statistical and systematic errors summed 
in quadrature. Right panel: Elliptic flow $v_2$ for pions, kaons and protons (solid lines) for the same parameters, together with ALICE data (circles). We also show the elliptic flow calculated in the ideal case, i.e.\ without the shear stress term $\delta f$ 
in the particle distribution (dashed lines). Ratios of calculations to dtata are shown below the panels.}
\end{figure}

Using the preferred parameters, we calculate the transverse momentum spectra and elliptic flow $v_2$ for the 30-40\% ALICE centrality bin. We show these calculations together with the data in Fig.\ \ref{fig:3040data}. The bottom of the figure shows the ratio of calculation over 
data. For the majority of $P_T$-bins the deviation is less than 5\%, and it rarely exceeds 20\%. If the experimental
error bars are included, the ratio is consistent with one almost everywhere in the RFR.

\begin{table}[t]
\centering
\begin{tabular}{|l||l|l|l|l|l|l|l|}
 \hline
Centrality & $\tau$ (fm/$c$) & $T$ (MeV) & $\alpha_0/c$ & $n$ & $R_y/R_x$ & $\alpha_2/c$ & $4\pi \eta/s$ \\
\hline\hline
\multicolumn{8}{|l|}{ALICE 2.76 TeV }  \\
 \hline\hline
10-20\% & 14.76 & 113.3 & 0.856 & 0.78 & 1.143 & 0.0355 & 5.89 \\
 \hline
20-30\% & 13.05 & 118.0 & 0.839 & 0.80 & 1.200 & 0.0517 & 5.45 \\
 \hline
30-40\% & 11.41 & 121.9 & 0.830 & 0.87 & 1.270 & 0.0564 & 4.06 \\
 \hline
40-50\% & 9.96 & 125.5 & 0.835 & 1.07 & 1.362 & 0.0472 & 2.46 \\
 \hline
50-60\% & 8.72 & 130.1 & 0.823 & 1.27 & 1.433 & 0.0427 & 1.66 \\
 \hline\hline
\multicolumn{8}{|l|}{PHENIX 0.2 TeV }  \\
 \hline\hline
10-20\% & 10.9 & 121.2 & 0.734 & 0.80 & 1.090 & 0.0463 & 3.32 \\
 \hline
20-30\% & 9.28 & 123.5 & 0.742 & 0.94 & 1.167 & 0.0528 & 1.98 \\
 \hline
30-40\% & 9.08 & 124.2 & 0.733 & 0.90 & 1.227 & 0.0576 & 1.64 \\
 \hline
40-50\% & 7.15 & 132.2 & 0.704 & 1.03 & 1.312 & 0.0631 & 1.08 \\
 \hline
50-60\% & 6.96 & 135.3 & 0.689 & 1.00 & 1.354 & 0.0630 & 0.93 \\
\hline\hline
\multicolumn{8}{|l|}{MUSIC 0.2 TeV }  \\
 \hline\hline
$b=7.5$ fm	 & 8.0 & 131.9 & 0.768 & 0.9 & 1.220 & 0.0357 & 1.06 \\
 \hline
\end{tabular}
\caption{Preferred values for the parameter set $\mathcal P$ obtained for different centrality bins for ALICE and PHENIX data in the regular fit range. We also show the results of fits to MUSIC pseudo data as discussed in the text.}
\label{table:result}
\end{table}

We analyze other centrality bins of ALICE analogously. The results for all ALICE centrality bins are summarized in Tab.\ \ref{table:result}. 
We note that the general trends of parameters as functions of centrality are consistent with expectations. 
The freeze-out temperature $\tfo$ rises toward smaller systems. The boundary velocity $\alpha_0$ reduces slightly at the same time.
The (spatially) averaged radial velocity (not shown) drops more significantly due to the concurrent change in the radial shape 
parameter $n$. 
These systematic trends give an important qualitative check of the fit results. However, we will not be interested in further interpretation of
fit parameters other than the temperature and specific shear viscosity.
The ALICE data sets provide us with a range of temperatures from roughly 113 MeV to 130 MeV.

Generally, the azimuthal flow deformation parameter $\alpha_2$ and spatial deformation $R_y/R_x$ as well as the specific shear viscosity $\eta/s$ are most sensitive to the elliptic flow data. 
We indicate the sensitivity of the calculated elliptic flow on $\eta/s$ at freeze-out by also showing in Fig.\ \ref{fig:3040data} the elliptic flow computed with the same parameters but without the correction term $\delta f$. As expected, at large $P_T$ the corrections from 
$\delta f$ are largest, thus extracted values of $\eta/s$ are very sensitive to $v_2$ at large $P_T$. 
Note however that despite the $p^2$-dependence of $\delta f$ in the local rest frame,
 the correction to $v_2$ due to $\delta f$ does not have to strictly vanish at small transverse momenta $P_T$ in the 
\emph{lab frame}. 
We have to be mindful that $\delta f$ can not be too large. As discussed earlier, higher order corrections in shear stress would have to be taken into account if $\delta f \approx f$. We have chosen the RFR such that $v_2$ starts to deviate from the equilibrium behavior at large $P_T$, but we generally exclude points for which the slope of $v_2$ turns negative. In the RFR we find that the
viscous correction is largest for protons, topping out at 19\% for the largest $P_T$-bin in the spectrum for the 40-50\% centrality bin.
For kaons and pions the largest corrections for the spectra we find are 11\% and 4\%, respectively. The typical size of viscous 
corrections is much smaller than the maximum numbers quoted here.

\begin{figure}[tb]
\centering
\includegraphics[width=3.5 in]{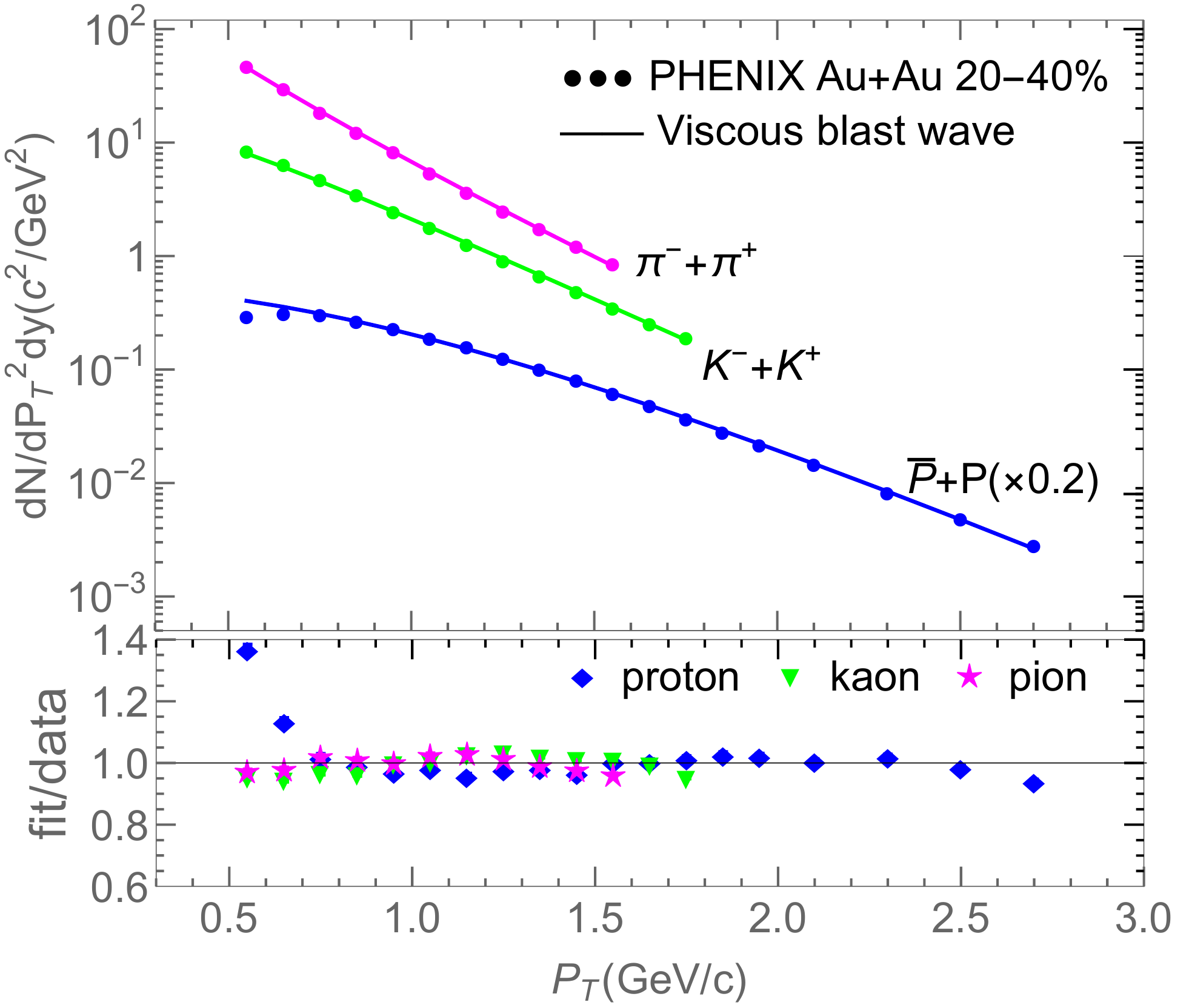}
\includegraphics[width=3.5 in]{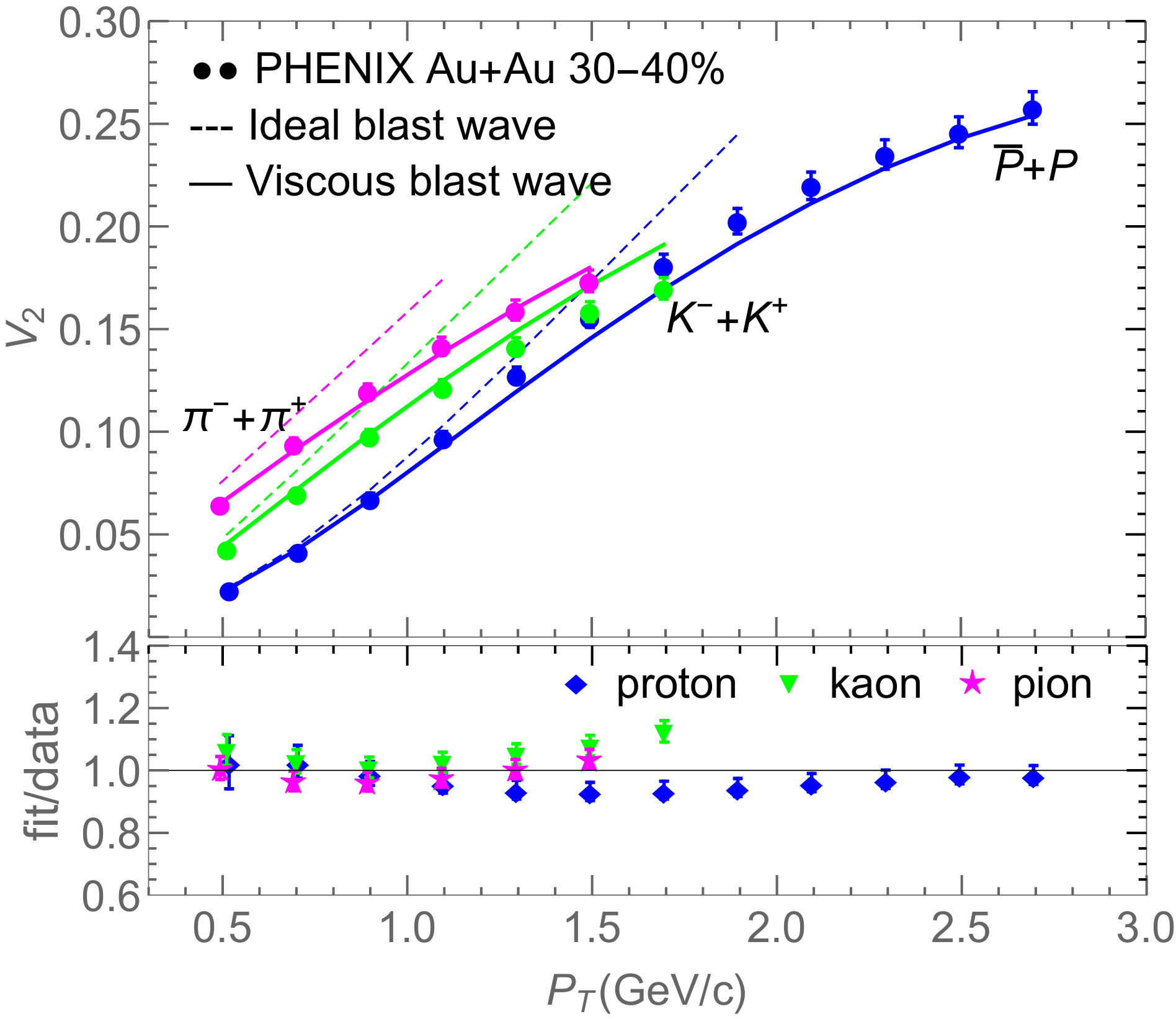}
\caption{\label{fig:3040phenix}
Same as Fig.\ \ref{fig:3040data} for the PHENIX 30-40 $\%$ centrality bin (20-40\% for the spectrum). Statistical errors only are shown for PHENIX spectra. }
\end{figure}

\begin{figure}[tb]
\centering
\includegraphics[width=4.5 in]{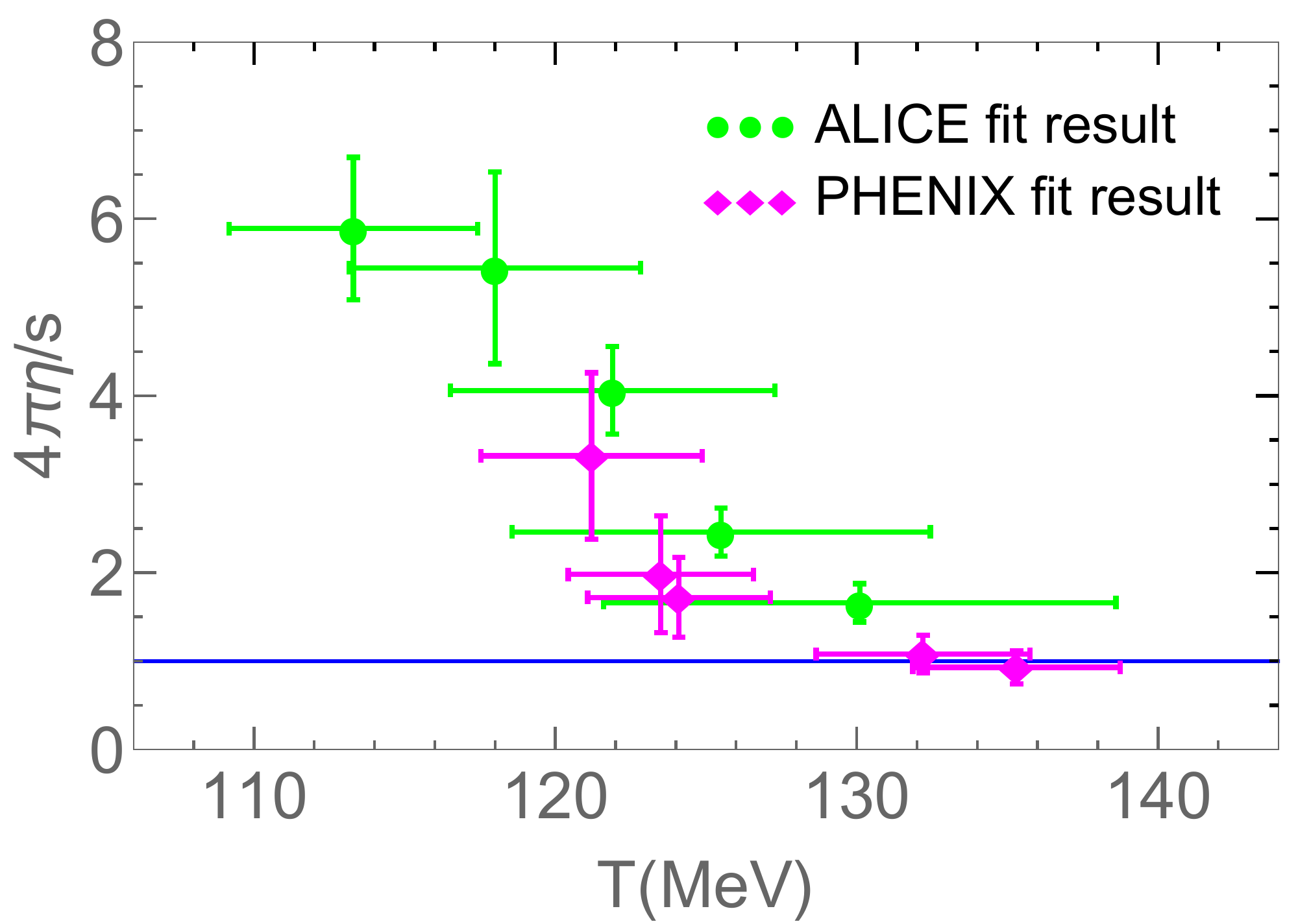}
\caption{Specific shear viscosity $\eta/s$ at corresponding kinetic freeze-out temperature $T$ extracted from the available ALICE and 
PHENIX centrality bins. Uncertainties shown are combined uncertainties of type (III) and (IV) explained in the text. Note that the 
values of the chemical potentials for stable hadrons are non-zero for all of these points.}
\label{fig:etas}
\end{figure}

We repeat the analysis with data from PHENIX in 200 GeV Au+Au collisions. The preferred, average values are also summarized in 
Tab.\ \ref{table:result}. The fits with preferred parameter values for one centrality bin are shown in Fig.\ \ref{fig:3040phenix} together with PHENIX data. 
The behavior of parameters as a function of centrality is similar to the one discussed for the ALICE data sets. The extracted 
temperatues range, roughly 122 MeV to 136 MeV overlaps with ALICE. It is an important consistency check that the 
extracted values for $\eta/s$ are consistent between ALICE data taken at $\sqrt{s_{NN}} = 2.76$ TeV and PHENIX data taken 
at $\sqrt{s_{NN}} = 0.2$ TeV, within uncertainties.
We summarize the results for $\eta/s$ vs temperature $T$ from all data sets in Fig.\ \ref{fig:etas}. The main qualitative
feature is a decrease in $\eta/s$ with increasing temperature, as would be expected from general principles. However, values close to
the lower bound are already reached at the upper end of the temperature range.

Let us now turn to a discussion of the uncertainties in our analysis. We can group them into four categories, ranging from 
basic statistical errors to rather fundamental in nature: (I) Fundamental limitations in 
our freeze-out ansatz that are shared between blastwave and fluid dynamics, e.g.\ from the validity of the Navier-Stokes approximation, and the assumption of a sharp freeze-out hypersurface. (II) Uncertainties and biases from assumptions made specifically in our blastwave model, e.g.\ the simple ansatz for the freeze-out hypersurface and the flow field, and the lack of resonance decays and bulk stress effects. (III) Uncertainties from our choice of external parameters and choice of fit ranges. (IV) Uncertainties from the errors in experimental data and the quality of the Gaussian
emulator. A thorough analysis of item (I) is beyond the scope of this paper and can not be achieved within the blastwave model. However we will attempt to analyze the other three sources of uncertainty.

\begin{table}[tb]
\begin{tabular}{|l|l|l|l|l|l|l|}
 \hline
Fit range (GeV/c)& proton & kaon & pion & $T$ (MeV) & $4\pi\eta/s $ \\ 
\hline
Low (LFR) & 0.325-2.05 & 0.225-1.25 & 0.19-0.825 & 113.4 & 3.85 \\ 
 \hline
Regular (RFR) & 0.325-3.1 & 0.225-2.25 & 0.525-1.65 & 121.9 & 4.06 \\ 
 \hline
High (HFR) & 1.25-3.1 & 0.725-2.25 & 0.825-1.65 & 125.2 & 3.43 \\ 
 \hline
\end{tabular}
\caption{\label{table:range} Definitions of different fit ranges for ALICE spectrum data in the 30-40 $\%$ centrality bin. The ranges for
  $v_2$ data are chosen commensurately. Also shown are the extracted temperature and specific shear viscosity for each fit range.} 
\end{table}

Uncertainties in extracted parameters from the error bars in our data sets and statistical analysis (type IV), are provided by the MADAI code. We quote the widths $\sigma_T^{\mathrm{stat}}$, $\sigma_\eta^{\mathrm{stat}}$ of temperature and specific shear viscosity for each centraliy bin and energy. We estimate uncertainties summarized under (III) by systematically
varying the underlying assumptions. E.g., as discussed earlier we choose alternative fit ranges which are shifted to lower (LFR) or larger (HFR) $P_T$. Limitations apply as we do not want to push too far into regions where we expect our blastwave to fail, see the discussion of fit ranges in Sec.\ \ref{sec:3}.
We discuss results once more for the 30-40 $\%$ ALICE centrality bin as an example. For the uncertainty analysis we 
focus on the results for the extracted temperature and specific shear viscosity. Table \ref{table:range} shows the three fit ranges, LFR, RFR, HFR for all three particle
species for this data set. Both temperature and $\eta/s$ show moderate dependencies on the fit range. This is expected for the temperature, where a change in $P_T$ samples different admixtures of resonance decays in spectra with different slopes and thus apparent temperatures.
We parameterize the deviations seen from the RFR values as Gaussian fluctuations with widths $\sigma_k^{\mathrm{range}}$ ($k=T, \eta$).
We repeat this analysis for all other centralities and energies with qualitatively similar results.

\begin{table}[bt]
\centering
\begin{tabular}{|l|l|l|l|l|}
 \hline
& $\mu_\pi$ (MeV) & T (MeV) & $4\pi\eta/s$ \\ 
 \hline
less & 46 & 121.0 & 4.01 \\ 
 \hline
regular & 61 & 121.9 & 4.06 \\ 
 \hline
more & 76 & 122.7 & 3.85 \\ 
 \hline
\end{tabular}
\caption{The freeze-out temperature $T$ and specific shear viscosity $\eta/s$ extracted for different values of pion chemical potential 
$\mu_\pi$ as explained in the text, for the ALICE 30-40$\%$ centrality bin.
\label{tab:chempi}}
\end{table}

\begin{table}[bt]
\centering
\begin{tabular}{|l|l|l|l|l|}
 \hline
& $c_s^2 (c^2)$ & $T$ (MeV) & $4\pi \eta/s$ \\ 
 \hline
small & 0.15 & 121.8 & 4.27 \\ 
\hline
regular & 0.166 & 121.9 & 4.06 \\ 
 \hline
large & 0.182 & 122.0 & 3.85 \\ 
 \hline
\end{tabular}
\caption{The same as Tab.\ \ref{tab:chempi} for a variation of the speed of sound squared $c_s^2$ for the ALICE 30-40$\%$ centrality bin.
\label{tab:cs2}}
\end{table}

\begin{table}[bt]
\centering
\begin{tabular}{|l|l|l|l|l|}
 \hline
& $c_\tau$ & $T$ (MeV) & $4\pi \eta/s$ \\ 
 \hline
small & 0.666 & 121.6 & 3.82 \\ 
 \hline
regular & 0.720 & 121.9 & 4.06 \\ 
 \hline
large & 0.781 & 121.2 & 4.12 \\ 
 \hline
\end{tabular}
\caption{The same as Tab.\ \ref{tab:chempi} for a variation of time-averaged surface velocity parameter $c_\tau$ for the ALICE 30-40$\%$ centrality bin.
\label{tab:ctau}}
\end{table}

\begin{table}[t]
\centering
\label{tab:uncert}
\begin{tabular}{|l|l|l|l|l|l|l|}
 \hline
 Origin of uncertainty & Stat.\ analysis & fit range & $\mu_\pi$ & $c_s^2$ &$c_\tau$ & total $\sigma$ \\
 \hline
$T$ (MeV) & 1.90 & 4.97 & 0.69 & 0.08 & 0.29 &  5.38 \\
 \hline
$4\pi\eta/s$ & 0.35 & 0.26 &0.10 & 0.17 & 0.13 &  0.50 \\

 \hline
\end{tabular}
\caption{A summary of uncertainties $\sigma_T^i$ and $\sigma_\eta^i$ for temperature and specific shear viscosity, respectively, 
 for the 30-40\% ALICE centrality bin. Here $i$ refers to the different contributions discussed in the text.}
\end{table}

As discussed earlier we also study the effects of variations in the chemical potential, speed of sound squared, and the 
expansion parameter $c_\tau$. Table \ref{tab:chempi} shows the values for $T$ and $\eta/s$ extracted for the 30-40\%
ALICE centrality bin for $\pm 15$ MeV variations in the pion chemical potential. We find that the temperature is rather insensitive
to variations of $\mu_\pi$ while $\eta/s$ displays moderate sensitivity. We again assign Gaussian widths $\sigma_k^\mu$ 
($k=T, \eta$) for the uncertainty from this source. We proceed similarly with variations in $c_s^2$, see Tab.\ \ref{tab:cs2} and $c_\tau$ (Tab.\ \ref{tab:ctau}). In both cases 
we find again very little influence on the extracted temperature. Finally we combine the uncertainties of types (III) and (IV) by adding the individual widths $\sigma^i_T$ and $\sigma^i_\eta$ in quadrature. Note that this assumption of Gaussian behavior here is simply an approximation. The error bars in $T$ and $\eta/s$ shown in Fig.\ \ref{fig:etas} are the result of this analysis. Table \ref{tab:uncert} summarizes the uncertainties for the ALICE 30-40\% centrality bin.

Finally we make an attempt to quantify uncertainties of type (II). To this end we use particle spectra and $v_2$ created 
with the viscous fluid dynamic code MUSIC \cite{Ryu:2015vwa}, and apply the same analysis that we have used for ALICE and PHENIX
data. In this case we know the precise temperature of freeze-out (set to 135 MeV) and the specific shear viscosity (set to $\eta/s=1/4\pi$ 
at freeze-out and throughout the evolution), and we can compare to the result of our analysis. We use an average (i.e. smooth) 
Au+Au collision system with impact parameter $b=7.5$ fm. The key difference between the fluid dynamic system at freeze-out and
the blastwave are the more realistic freeze-out hypersurface, more complex shape of the flow field, and the presence of both
bulk stress at freeze-out and hadronic decays after freeze-out in MUSIC. By comparing our extracted temperature and 
specific shear viscosity with the one set in the fluid dynamic system we can gain insights what deviations we have to expect through the absence of these features in the blastwave.

\begin{figure}[thpb]
\centering
\includegraphics[height=4.2 in,width=4.2 in]{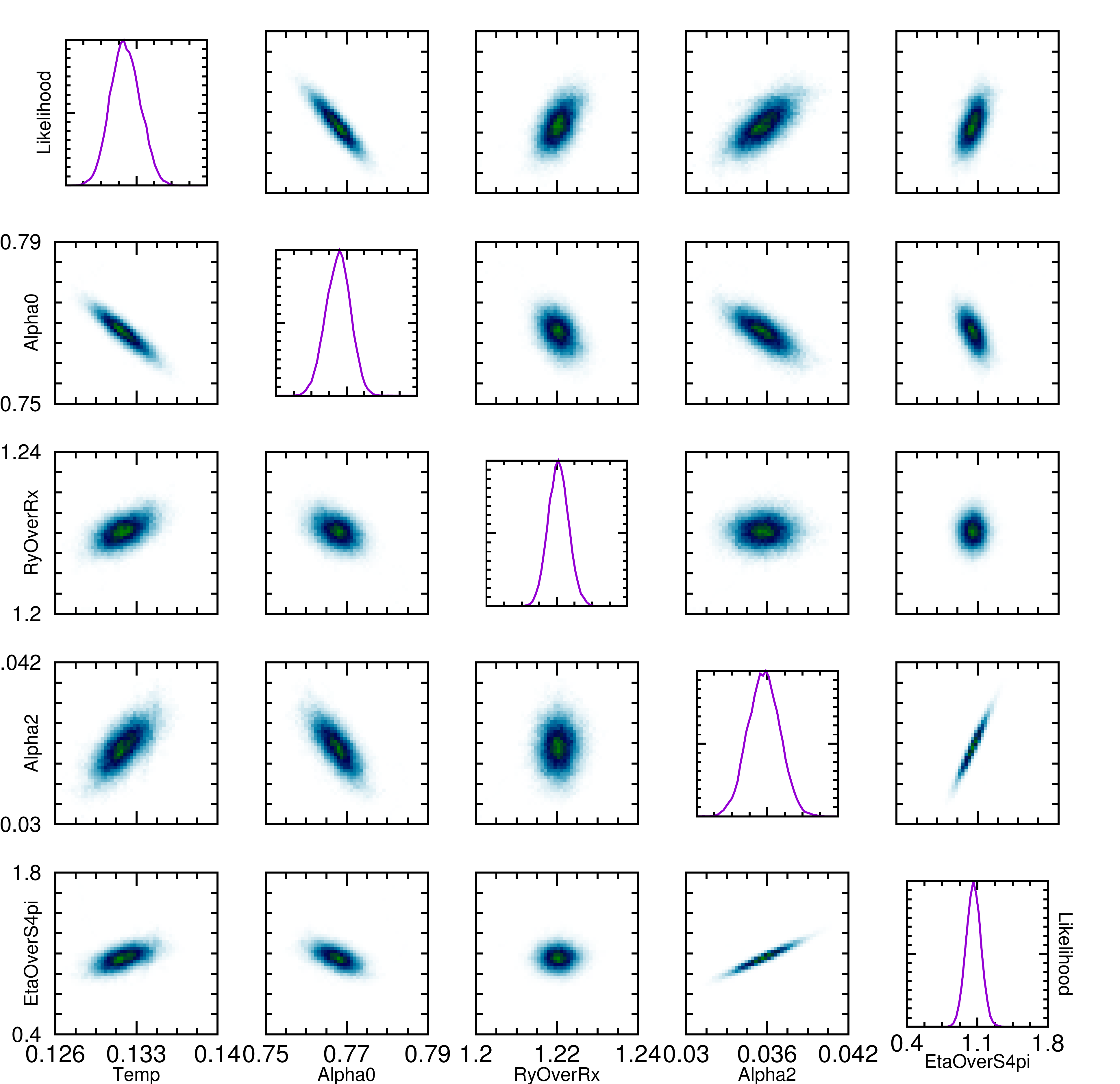}
\caption{\label{fig:hydromadai} Likelihood and correlations for the parameter set $\mathcal{P}$, reduced by $n$, for a fit
to MUSIC pseudo data.
}
\end{figure}

\begin{figure}[tbhp]
\centering
\includegraphics[width=3.5 in]{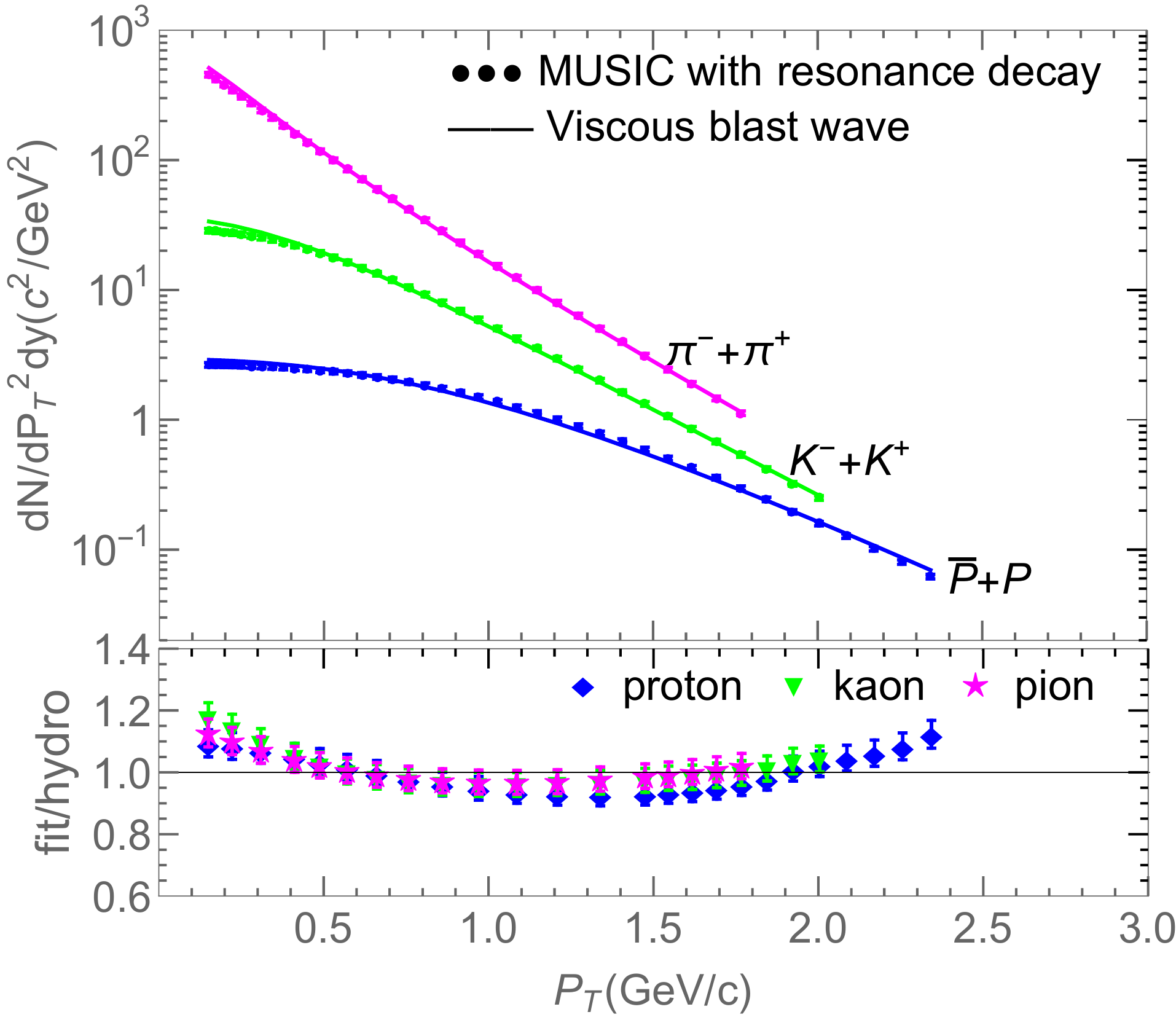}
\includegraphics[width=3.5 in]{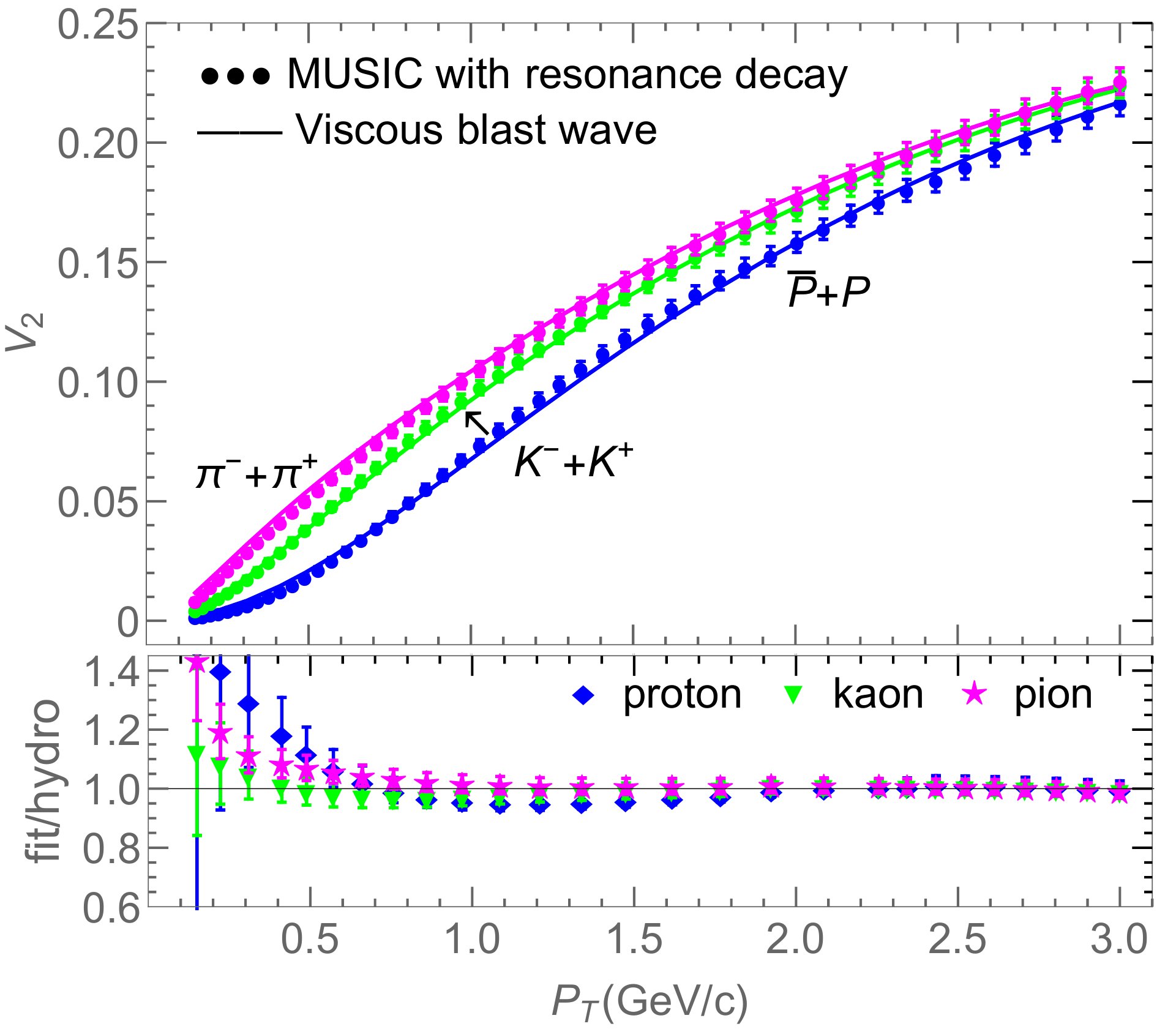}
\caption{\label{fig:hydro}
Same as Fig.\ \ref{fig:3040data} but for a fit to spectra and elliptic flow generated by the MUSIC code as described in the text.}
\end{figure}

We need to specify a fit range and error bars for the MUSIC pseudo data. We choose them to be roughly consistent with the fit ranges used
in the actual data analysis. These values can be found in Tables \ref{table:allrange} and \ref{table:error}.
The relative error of 2\% mentioned for $v_2$ in the latter table is supplemented by a pedestal of 0.001 which needs to 
be added to give data points very close to zero reasonably sized errors. 
In our analysis we find that the MUSIC pseudo data leads to a larger degree of degeneracy coming from correlations between 
fit parameters. We have made the choice to remove the parameter $n$ from the fit parameter tuple $\mathcal{P}$ and to set its value to
$n=0.9$, commensurate with values found from RHIC data for similar impact parameter. This reduction leads to crisply
defined fits for the remaining parameters. 
We show the likelihood and correlation plot for the MUSIC analysis in 
Fig.\ \ref{fig:hydromadai}, and the preferred values extracted at the bottom of Tab.\ \ref{table:result}. We find $T=131.9$ MeV
and $\eta/s = 1.06$. Fig.\ \ref{fig:hydro} shows calculations with our blastwave using the extracted parameters together with
the MUSIC pseudo data and assigned error bars.

$n$ was chosen to be removed from the fit as a somewhat obscure parameter that we have no further interest in. In fact such a shape parameter 
for the flow field is absent in many simpler blastwave parameterizations on the market. Nevertheless one might worry what this 
choice means for our uncertainty analysis. The same can be said about our choice or error bars for the pseudo data.
We have varied both $n$ and the error bars of the MUSIC pseudo data. We find that the extracted value of the temperature is rather
stable, at about $3$ MeV below the value set in MUSIC. The extracted value for the specific shear viscosity can change by up to 30\%, for reasonable variations of $n$.
In conclusion, our exploration of uncertainties of type (II) for this data set suggest that temperatures extracted with the blastwave are likely biased to find lower than actual temperatures, but this bias is only about 3 MeV. There is also an additional uncertainly on $\eta/s$ of up to 30\% which could be added to the total tally of uncertainties. Note again
that Fig.\ \ref{fig:etas} only shows the uncertainty in $T$ and $\eta/s$
from type (III) and (IV). 
Our analysis of type (II) uncertainties could certainly be made more quantitative. In principle we have to run the same analysis described here for every ($T$-$\eta/s$) point that we have extracted from data. We keep this discussion for a future publication.

\section{Discussion}
\label{sec:discussion}

\begin{figure}[tb]
\centering
\includegraphics[width=3.5 in]{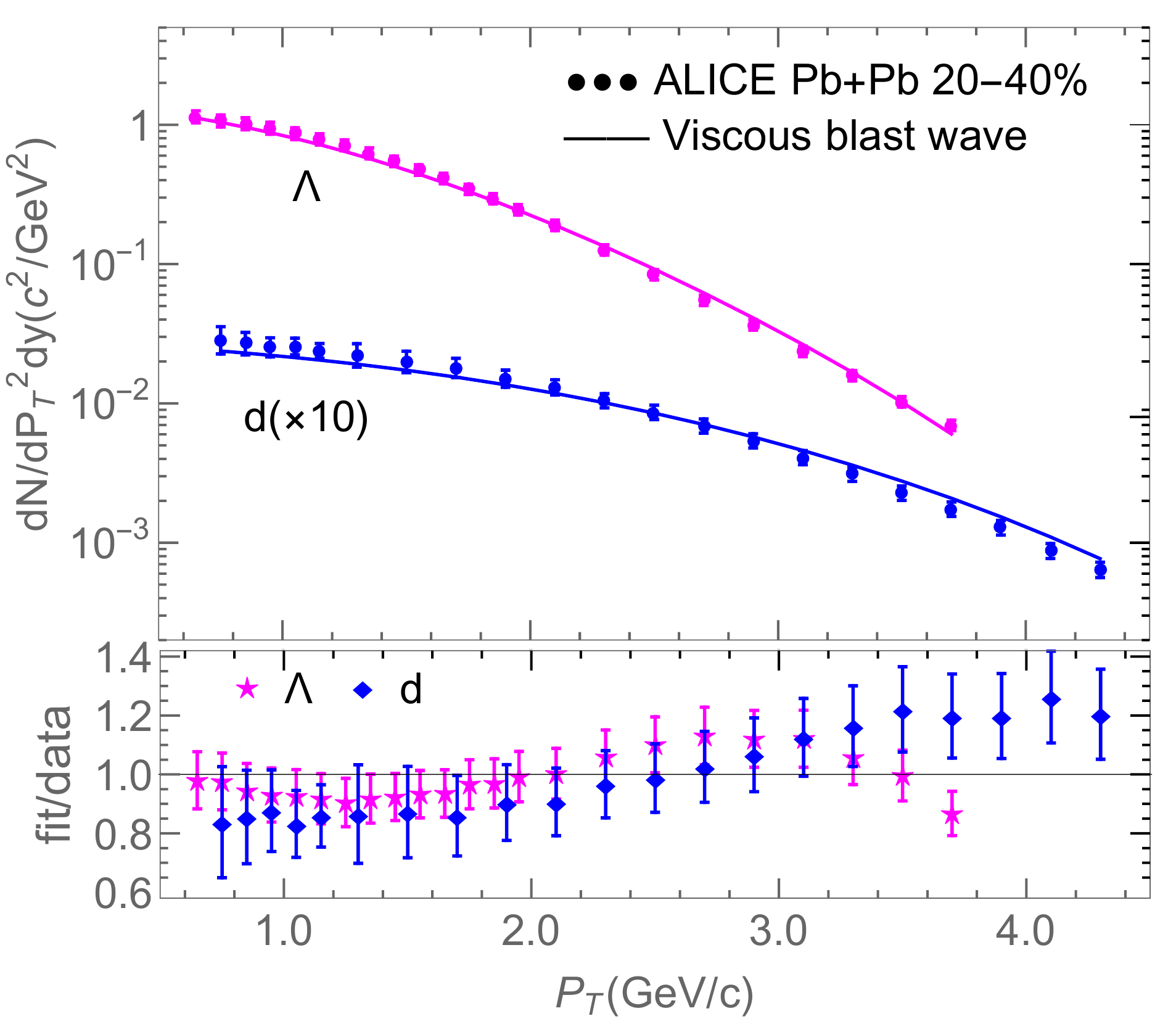}
\includegraphics[width=3.5 in]{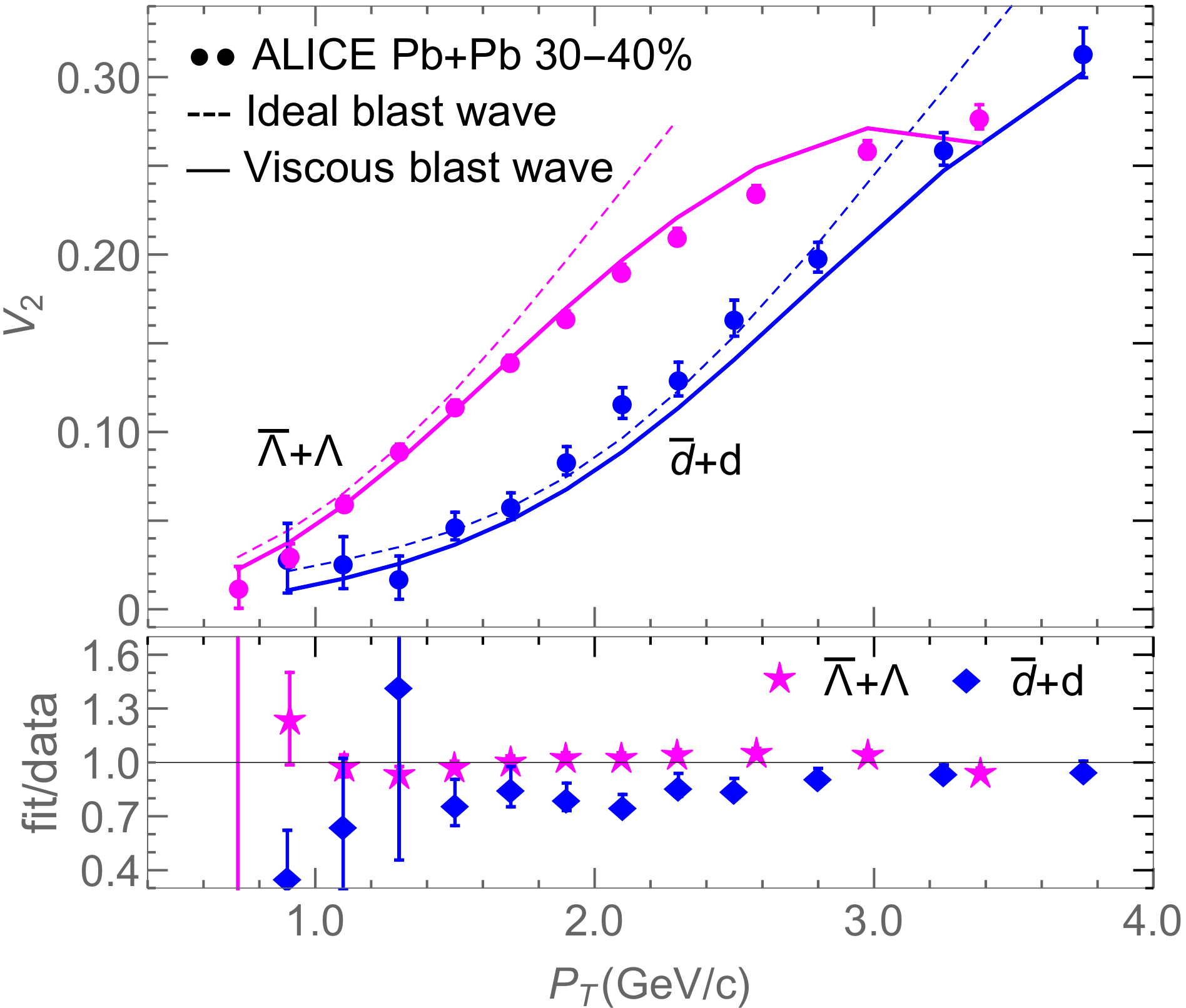}
\caption{\label{fig:3040prediction}
Left panel: Transverse momentum spectra for $\Lambda$s and deuterons (solid lines), respectively, calculated for the
20-40 $\%$ centrality bin in Pb+Pb collisions together with ALICE data (symbols). Right panel: Elliptic flow $v_2$ for $\Lambda+\bar\Lambda$ and $d+\bar d$ (solid lines) in the 30-40\% centrality bin together with ALICE data (circles). We agin show the elliptic flow 
calculated in the ideal case as well. In both cases the preferred parameters for the 30-40\% centrality bin extracted for stable
charged hadrons has been used.}
\end{figure}

We have introduced a blastwave model with viscous corrections due to shear stress in the Navier-Stokes approximation.
The blastwave model can obtain excellent fits to hadron spectra and $v_2$ over a large range of $P_T$. The viscous correction
term helps to describe the slow down of the growth of $v_2$ with $P_T$. This model provides a reliable instrument that can give useful snapshots of the dynamically evolving fireball.

To further demonstrate the usefulness we plot predictions for the spectra and $v_2$ for two more particles, the $\Lambda$ baryon and the deuteron $d$, in a mid-central bin as examples. The results are shown in Fig.\ \ref{fig:3040prediction} together with
ALICE data \cite{Acharya:2017dmc,ABELEV:2013zaa,Abelev:2014pua}. Note that our calculation is a prediction in the sense that  $\Lambda$ and deuteron data have not been used to fix the blastwave parameters. Chemical potentials for both species have been fixed to 344 and 314 MeV respectively. We find overall good agreement for this centrality bin.
This is interesting since there have been questions in both cases about the validity of a common freeze-out with stable hadrons. In 
particular the deuteron is often thought to be emerging from coalescence processes after freeze-out 
\cite{Acharya:2017dmc,Adamczyk:2016gfs,Zhu:2017zlb}. We find that, whatever the detailed mechnism of deuteron creation, the 
spectra and elliptic flow are described reasonably well by the same temperature and flow field that also describes stable hadrons, at least in 
mid-central collisions. 

\begin{figure}[tb]
\centering
\includegraphics[width=3.5 in]{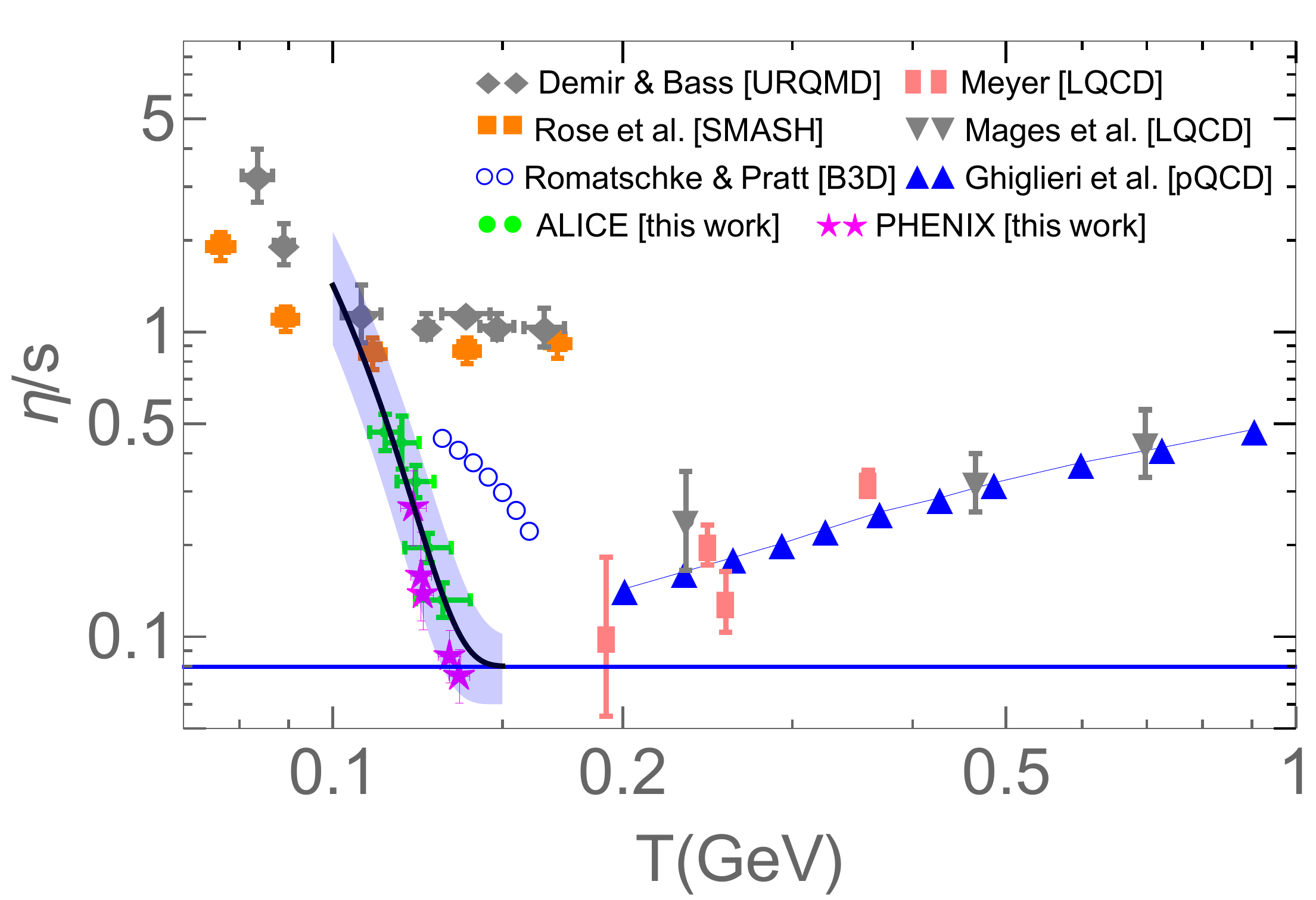}
\caption{\label{fig:final}
The results of this paper compared to various calculations of the specific shear viscosity $\eta/s$ as a function of temperature. 
A line and uncertainty band have been drawn through our points to guide the eye. Details in the text.
}
\end{figure}

Let us now turn to a discussion of the particular application of our blastwave we have focused on here.
From two different collision systems, Pb+Pb at LHC energy and Au+Au at top RHIC energy we have extracted
several $\eta/s$-vs-$T$ points that are consistent with each other within estimated uncertainties. They give us a band,
indicated in the figure, that roughly stretches from 110 MeV to 140 MeV. The center line of the band can be approximately
parameterized as $\eta/s = 1.47\times 10^5\times(0.155 - T)^4 + 0.08$ where $T$ is measured in GeV.
Note again that uncertainties of type (II) are not included here. The bias discussed at the end of the last section indicates that 
points might have to be shifted up in temperature by about 3 MeV. 
We also show results for the hadronic phase from hadronic cascades URQMD \cite{Demir:2008tr}, B3D \cite{Romatschke:2014gna} and SMASH \cite{Rose:2017bjz}. 
They generally show larger values of $\eta/s$ above $T\sim100$ MeV.
One could speculate that below $T\sim100$ MeV the results might converge within uncertainties, as the URQMD and SMASH results switch 
their behavior to a temperature slope similar to our results. Unfortunately we do not have the data points to confirm this.
We also show several calculations of the specific shear viscosity in the QGP phase, from lattice QCD \cite{Meyer:2007ic,Meyer:2009jp,Mages:2015rea} and using next-to-leading perturbative QCD \cite{Ghiglieri:2018dib}.
Overall these results together are consistent with the idea of a minimum of the specific shear viscosity around the pseudocritical 
temperature $T_c$. Our result specifically would indicate a rather broad minimum where interactions in the hadronic phase continue
to be strong just below $T_c$ while hadronic transport suggests a more abrupt change below $T_c$. 

However, we need to keep in mind that relatively large chemical potentials for stable hadrons build up in the collision systems that
we have analyzed here. E.g.\ the chemical potential for pions is as large as 70 MeV at the lowest temperature points we have extracted. Thus
Fig.\ \ref{fig:final} is a projection of a more complicated plot with additional chemical potential axes. Studies with hadronic transport
have indicated that finite chemical potentials can indeed lead to smaller values of $\eta/s$ \cite{Demir:2008tr} in this picture.

The fate of $\eta/s$ in the hadronic phase continues to be intriguing. We have added a scenario, based on extraction from data, that predicts a 
steep rise of $\eta/s$ while the temperature drops from 140 and 110 MeV and chemical potentials increase.
Our approach is rooted in data taken in heavy ion collisions but has built in uncertainties. We have quantified the more accessible
uncertainties (IV) and (III) related to the analysis itself and to systemic uncertainties from choices made during the analysis. We have also
made a first attempt to estimate the weaknesses of the blastwave compared to a full fluid dynamic simulation, i.e.\ uncertainties of type (II).
More fundamental uncertainties remain which may be quantified elsewhere.
Certain aspects of the current analysis will be improved in the near term future. For example the detailed energy dependence of the 
shear stress term, parameterized by $\lambda$, and the effects of bulk stress could be included, albeit at the expense of adding two parameters
to the analysis. One could also include an analysis of the asymmetry coefficient $v_4$, which requires a generalization
of both hypersurface and flow field of the blastwave. Lastly, resonances and their decays could in principle be included in the calculation.

\begin{acknowledgments}
ZY would like to thank Yifeng Sun for useful discussions. RJF would like to thank Charles Gale and Sangyong Jeon for comments and their hospitality 
at McGill University where part of this work was carried out. We thank Mayank Singh for providing MUSIC support.
RJF also acknowledges useful discussions with Amaresh Jaiswal and Ron Belmont. We thank Che-Ming Ko, J\"urgen Schukraft and Ulrich Heinz for valuable comments.
This work was supported by the US National Science Foundation under award nos.\ 1516590, 1550221 and 1812431.
\end{acknowledgments}


\end{document}